%% file: main.tex
\documentclass[journal]{IEEEtran}

\usepackage[english]{babel}
\usepackage[autostyle]{csquotes}
\MakeOuterQuote{"}
\usepackage{flushend}

\usepackage{booktabs} 
\usepackage{colortbl}
\usepackage{xcolor}

\usepackage{cite}
\usepackage[hyphens]{url}
\usepackage{amssymb,amsmath,amsthm, latexsym}

\usepackage[frozencache,cachedir=.]{minted}
\usepackage{enumerate}
\mathchardef\mhyphen="2D
\usepackage{algpseudocode}
\usepackage[ruled,vlined,linesnumbered]{algorithm2e}
\SetAlCapNameFnt{\small}
\SetAlCapFnt{\small}
\SetAlgorithmName{Alg.}{Alg.}{List of Algorithms}

\usepackage{graphicx}
    \graphicspath{{./img/}}
\usepackage{float}

\addto\captionsenglish{\renewcommand{\figurename}{Fig.}}

\newcommand{\squeeze}{\vspace{-2.0mm}}

\usepackage[font=scriptsize]{caption}
\begin{document}

\title{GDPR-Compliant Personal Data Management: A Blockchain-based Solution}

\author{Nguyen~Binh~Truong,~\IEEEmembership{Member,~IEEE,}
        Kai~Sun,~\IEEEmembership{Senior Member,~IEEE,}
				Gyu~Myoung~Lee,~\IEEEmembership{Senior Member,~IEEE,}	
        and~Yike~Guo,~\IEEEmembership{Fellow,~IEEE}
\thanks{N.B. Truong, K. Sun and Y. Guo are with Data Science Institute, Department of Computing, Imperial College London, London,
SW7 2AZ United Kingdom. E-mail: n.truong@imperial.ac.uk, k.sun@imperial.ac.uk, y.guo@imperial.ac.uk}
\thanks{G.M. Lee is with Department of Computer Science, Liverpool John Moores University, Liverpool,
L3 3AF United Kingdom. E-mail:g.m.lee@ljmu.ac.uk}
}

\markboth{IEEE Transaction on Information Forensics and Security, October~2019}%
{Truong \MakeLowercase{\textit{et al.}}: GDPR-Compliant Personal Data Management: A Blockchain-based Solution}

\maketitle

\input{0Abstract}


\IEEEpeerreviewmaketitle

\input{1Intro}
\input{2RelatedWork}
\input{3Challenges}
\input{4DesignConcept}
\input{5Deployment}
\input{6Analysis}
\input{7Conclusion}

\section*{Acknowledgment}
This research was supported by the HNA Research Centre for Future Data Ecosystems at Imperial College London.\squeeze


\bibliographystyle{IEEEtran}
\bibliography{refs}

\begin{IEEEbiography}[{\includegraphics[width=1in,height=1.25in,clip,keepaspectratio]{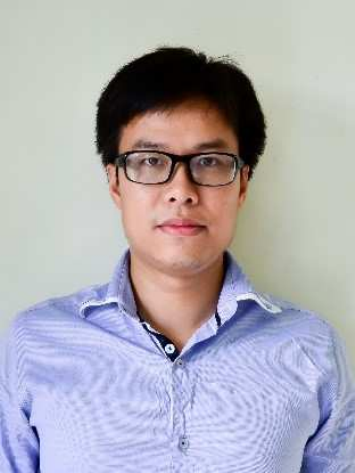}}]{Nguyen Binh Truong}
Dr. Nguyen B.Truong is currently a Research Associate at Data Science Institute, Department of Computing, Imperial College London, United Kingdom. He received his PhD, MSc, and BSc degrees from Liverpool John Moores University, United Kingdom, Pohang University of Science and Technology, Korea, and Hanoi University of Science and Technology, Vietnam in 2018, 2013, and 2008, respectively. He was a Software Engineer at DASAN Networks, a leading company on Networking Products and Services in South Korea from 2012 to 2015. His research interest is including, but not limited to, Security, Privacy and Trust for IoT, Blockchain, Personal Data Management, Fog, Edge and Cloud Computing.
\end{IEEEbiography}
\begin{IEEEbiography}[{\includegraphics[width=1in,height=1.25in,clip,keepaspectratio]{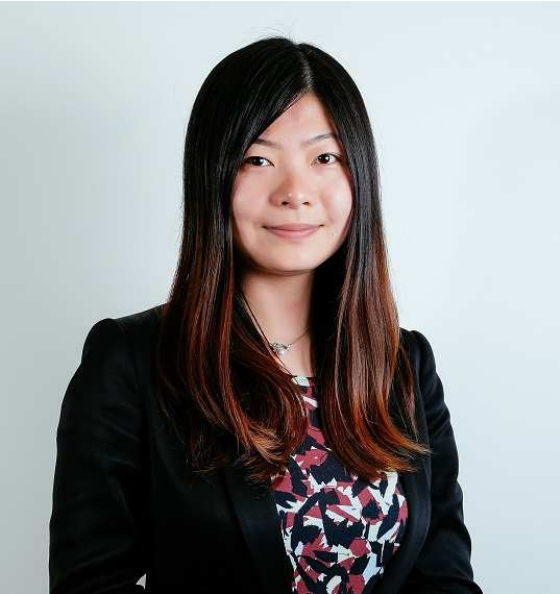}}]{Kai Sun}
Dr. Kai Sun received the BEng degrees in Computer Science from Harbin Institute of Technology and the University of Birmingham in 2009. She received the MSc degree and the PhD degree in Computing from Imperial College London, in 2010 and 2014, respectively. From 2014 to 2017, she was a Research Associate at the Data Science Institute at Imperial College London. She is currently the lab manager of the HNA Centre of Future Data Ecosystem. Her research interests include translational research management, network analysis and decentralised systems. 
\end{IEEEbiography}
\begin{IEEEbiography}[{\includegraphics[width=1in,height=1.25in,clip,keepaspectratio]{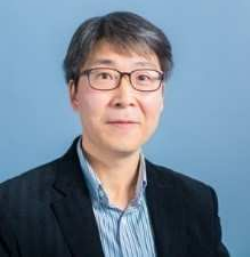}}]{Gyu Myoung Lee}
Dr. Gyu Myoung Lee received his BS degree from Hong Ik University and MS, and PhD degrees from the Korea Advanced Institute of Science and Technology (KAIST), Korea, in 1999, 2000 and 2007, respectively. He is currently a Reader at Department of Computer Science, Liverpool John Moores University, UK. He is also with KAIST as an adjunct professor. His research interests include Future Networks, IoT, and multimedia services. He has actively contributed to standardization in ITU-T as a Rapporteur, oneM2M and IETF. He is chair of the ITU-T Focus Group on data processing and management to support IoT and Smart Cities \& Communities.
\end{IEEEbiography}
\begin{IEEEbiography}[{\includegraphics[width=1in,height=1.25in,clip,keepaspectratio]{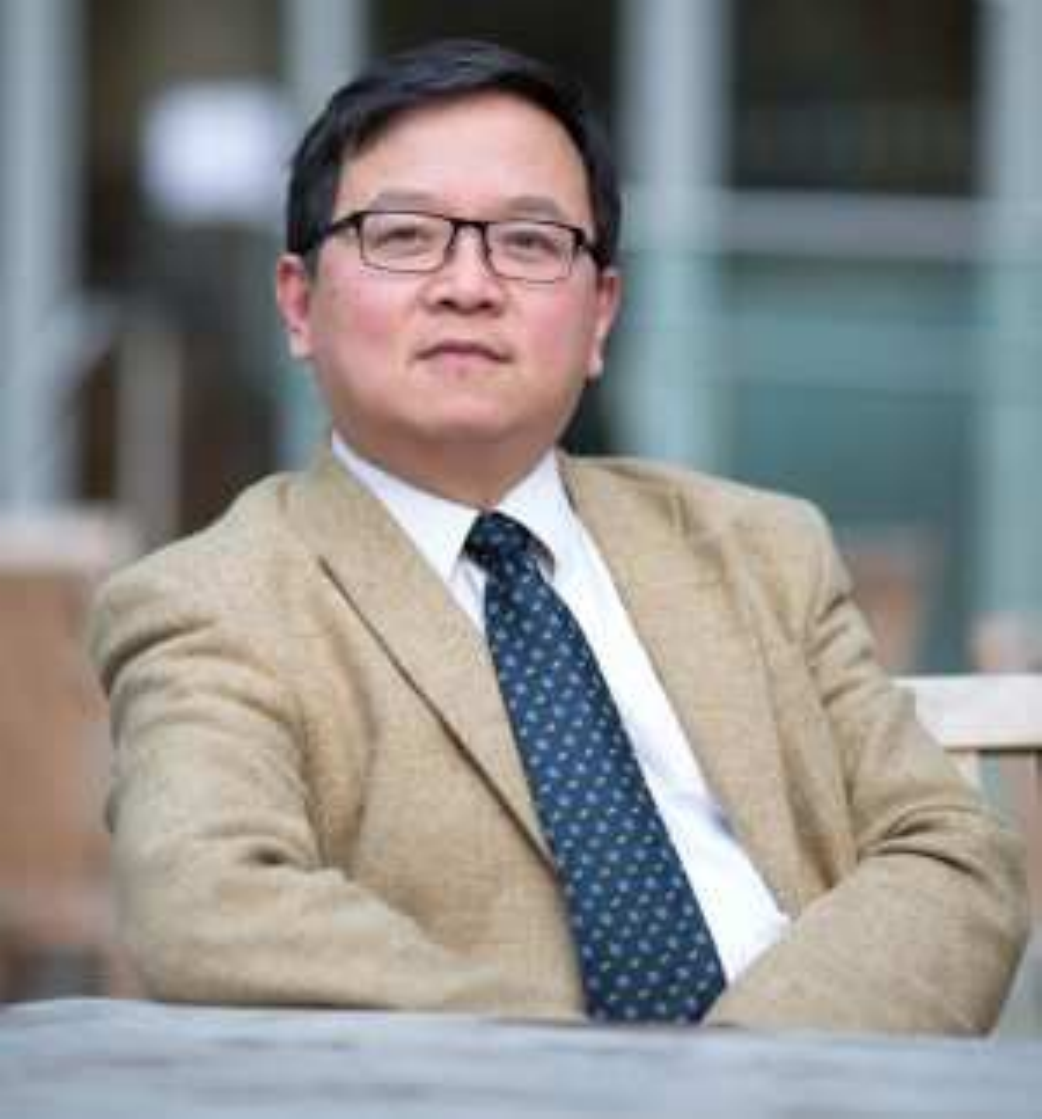}}]{Yike Guo}
Dr. Yike Guo (FREng, MAE) received the BSc degree in Computing Science from Tsinghua University, China, in 1985 and received the PhD in Computational Logic from Imperial College London in 1993. He is a Professor of Computing Science in the Department of Computing at Imperial College London, as well as the founding Director of the Data Science Institute at Imperial College. He is a fellow of the Royal Academy of Engineering. His research interests are in the areas of data mining for large-scale scientific applications including distributed data mining methods, machine learning and informatics systems.
\end{IEEEbiography}

\end{document}

%% file: 0Abstract.tex
\begin{abstract} \label{Abstract}
The General Data Protection Regulation (GDPR) gives control of personal data back to the owners by appointing higher requirements and obligations on service providers who manage and process personal data. As the verification of GDPR-compliance, handled by a supervisory authority, is irregularly conducted; it is challenging to be certified that a service provider has been continuously adhering to the GDPR. Furthermore, it is beyond the data owner's capability to perceive whether a service provider complies with the GDPR and effectively protects her personal data. This motivates us to envision a design concept for developing a GDPR-compliant personal data management platform leveraging the emerging blockchain and smart contract technologies. The goals of the platform are to provide decentralised mechanisms to both service providers and data owners for processing personal data; meanwhile, empower data provenance and transparency by leveraging advanced features of the blockchain technology. The platform enables data owners to impose data usage consent, ensures only designated parties can process personal data, and logs all data activities in an immutable distributed ledger using smart contract and cryptography techniques. By honestly participating in the platform, a service provider can be endorsed by the blockchain network that it is fully GDPR-compliant; otherwise, any violation is immutably recorded and is easily figured out by associated parties. We then demonstrate the feasibility and efficiency of the proposed design concept by developing a profile management platform implemented on top of the Hyperledger Fabric permissioned blockchain framework, following by valuable analysis and discussion.
\end{abstract}

%% file: 1Intro.tex
\section{Introduction} \label{2Intro}
\IEEEPARstart{T}{he} General Data Protection Regulation (GDPR) legislation came into force in May 2018 in all European Union (EU) countries. The GDPR is a major update to the data privacy regulations released in 1995, which is before the proliferation of cloud platforms and social media, let alone the scale of today's data usage. The provision of the GDPR is to ensure that personal data "can only be gathered legally, under strict conditions, for a legitimate purpose"; as well as to bring full control back to the data owners\footnote{https://gdpr-info.eu/}.

As the GDPR requirements are highly abstract, it is open to interpretation. In fact, each organisation has its own way to satisfy the new regulations; and to demonstrate the compliance. Supposedly, each EU member state provides a supervisory authority who is responsible for monitoring the GDPR-compliance. Organisations are required to demonstrate compliance only in case of suspicion of a violation or when a Data Subject (i.e., the owner of data, denoted as DS) lodges a complaint with the supervisory authority. In this regard, the challenge of complying with the GDPR is not because of lacking technical solutions for tackling down the GDPR requirements nor providing required mechanisms; it is because such solutions are designed and implemented under a centralised client-server architecture mindset. Due to the irregular verification of GDPR compliance, critical concerns on the lack of transparency have been imposed accordingly. In particular, it is unachievable for a Service Provider (SP) to prove that it has been continuously adhering to the GDPR using existing centralised solutions. Moreover, it is beyond the DS's capability to perceive whether an SP fully complies with the GDPR and effectively protects her data. For these reasons, GDPR-compliant personal data management is a well-suited scenario for the emerging blockchain technology (BC) to come into play. A BC platform implementing Smart Contracts (SCs) is expected to be a promising measure for these challenges thanks to its advanced features of decentralisation, transparency, tamper-resistance, and traceability.

Some research articles have stated potentials of the BC as a general-purpose data management and storage \cite{ref24, ref25, ref30, ref31, ref32, ref33, ref34, ref35, ref36, ref37, ref38}; however, they only provided preliminary methodological exploration or conceptual models without detailed technical analysis and implementation. In these articles, a holistic architecture of decoupling the BC, which is for accounting and auditing data access, from a storage layer, which physically stores data were adopted. Unfortunately, there are lacking of a comprehensive design concept and technical mechanisms to actualise the capability of the BC in personal data management and in complying with the GDPR requirements. In this article, we propose a design concept with technical mechanisms for a BC-based GDPR-compliant personal data management platform, along with a detailed implementation of the profile management system use-case built on top of a permissioned BC framework. The goal of the design concept is to preserve advanced features of BC and SCs in personal data management by leveraging distributed ledger and public-key cryptography technologies for complying with the manifold legal requirements of the GDPR \cite{ref01}. For this purpose, a BC network is designed to play as the roles of: \textit{(i)} a delegated authentication and authorisation server which is consolidated by a novel concept of decentralised \textit{access token}, \textit{(ii)} an automated access control manager, and \textit{(iii)} an immutable logging system a for parties who desire to access personal data stored in an off-chain Resource Server (RS).

By following the proposed design concept, a personal data management platform ensures that only designated DSs and Data Controllers (DCs) are permitted to create, update and withdraw consents; and only authorised Data Processors (DPs) can process personal data respecting rules defined in corresponding data usage policy agreed between the DSs and the DPs. The platform not only provides mechanisms for DS rights but also plays as a role of a DC for handling personal data processing and demonstrating data accountability. By honestly participating in the BC-based personal data management platform, an SP can be endorsed by the BC network that it is GDPR-compliant. Otherwise any violations are recorded in an immutable distributed ledger as a record of the infringements, which can be then used for the GDPR compliance investigation by supervisory authorities.

We demonstrate the feasibility and effectiveness of the proposed design concept by developing a system for managing personal profiles. The system, which is built on top of the Hyperledger Fabric (HLF) permissioned BC framework\footnote{https://www.hyperledger.org/projects/fabric} and cooperates with an honest RS for data storage, plays as a profile management service for a social networking SP. This system provides clients’ rights as well as facilitates the social networking SP's obligations, following by analysis and discussion on the GDPR-compliance, threat models and system performance. It is affirmed that the social networking SP is fully compliant with the GDPR requirements. We believe the proposed approach is a promising solution not only for GDPR-compliant personal data management but also for digital assets governance.

The rest of the article is organised as follows. Section II presents background and related work. Section III describes challenges and motivation. The design concept is proposed in Section IV following by the implementation of the profile management platform in Section V. Section VI provides the analysis and discussion about the platform. The last section concludes our work and outlines future research.

%% file: 2RelatedWork.tex
\section{Background and Related Work} \label{3RelatedWork}
In this section, relevant background knowledge on GDPR and BC and related work are presented. Table \ref{tb1} depicts some of the notions frequently used throughout this article.
\begin{table}[h]
	\centering
	\captionsetup{justification=centering}
	\caption{NOTATION TABLE WITH ENTRIES IN ALPHABETICAL ORDER}
	\label{tb1}
	\begin{tabular}{c|c}		
	    \toprule
		\textbf{Notation} & \textbf{Description} \\ [0.5ex]
		\midrule
		  \rowcolor{black!20} API & Application Programming Interface \\
		  BC & Blockchain \\
		  \rowcolor{black!20} BFT & Byzantine Fault Tolerance \\
		  C-ID & Complex Identity \\
		  \rowcolor{black!20} CA & Certificate Authority \\
		  CRUD & Create-Read-Update-Delete operations \\
		  \rowcolor{black!20} DBMS & Database Management System \\
		  DC & Data Controller \\
		  \rowcolor{black!20} DP & Data Processor \\
		  DS & Data Subject \\
		  \rowcolor{black!20} GDPR & General Data Protection Regulations \\
		  HLF & Hyperledger Fabric Blockchain framework\\
		  \rowcolor{black!20} IdM & Identity Management \\
		  MSP & Membership Service Provider \\
		  \rowcolor{black!20} OSN & Ordering Service Node \\
		  RS & Resource Server \\
		  \rowcolor{black!20} SC & Smart Contract \\
		  SP & Service Provider \\
			\rowcolor{black!20} TP & Third-Party \\
	\end{tabular} \\[0.1ex]
\end{table}

\subsection{The GDPR in a Nutshell}
The full GDPR are described in detail across $99$ articles covering all of the technical and admin principles around how commercial and public organisations process personal data \cite{ref_add_02}. GDPR lays out the means by which personal data is to be protected which are founded on a set of six core data processing principles: Lawfulness, Fairness, and Transparency; Purpose Limitation; Data Minimisation; Accuracy; Storage Limitation; Integrity and Confidentiality\footnote{https://gdpr-info.eu/art-5-gdpr/}. To preserve such principles, the GDPR clearly differentiates three roles (i.e., DS, DC and DP) and explicitly specifies associated rights and obligations under the EU data protection law. The goal of the GDPR legislation is to provide a DS full control over her personal data by specifying a variety of rights. The GDPR requires that personal data should be managed by a DC that assures the rights of the DS \cite{ref_add_02}. Such mechanisms enable the DS to impose consents and to arbitrarily withdraw the consents whenever needed. The DS is also able to trace back all activities on her data including who, what, why, when, and how the data is processed. Valid legal consents must be given by the DS to the DC for processing her personal data. The DC then takes appropriate measures to provide the rights of the DS; meanwhile determines the purposes for which and the method in which, the personal data is processed by DPs \cite{ref_add_01}.

Being compliant with the GDPR is not enough, DCs should also be able to demonstrate the compliance to supervisory authorities once required (when a supervisory authority has suspicion of a violation or when a DS lodges a complaint with the supervisory authority). In this case, the supervisory authority shall establish and make public a list of processing operations subjected to Data Protection Impact Assessment and the Privacy Impact Assessment requirements\footnote{https://gdpr-info.eu/issues/privacy-impact-assessment/}; then file a report of infringements if it is the case.

\subsection{Blockchain Technology}
The BC technology, indeed, is a set of diversified techniques including distributed systems, computer networks, databases, and cryptography playing as the role of a distributed ledger. The BC technology maintains a distributed immutable database constituted from a continuous growing list of blocks so-called a BC which records all transactions between entities in a network. In our article, the acronym \textit{BC} either refers to the technology or a specific chain-of-block database. By nature, a BC is inherently resistant to data modification. Once recorded, information in any given block cannot be altered retroactively as this would invalidate all hashes in the previous blocks in a BC; and break the consensus among nodes in the network. The concept of BC was introduced in Bitcoin in 2008 \cite{ref02}. Bitcoin is the first cryptocurrency that not only transacts digital currency in a secure manner but also resolves the long-standing problem of "double-spend" without the need for a trusted third-party. BC underpins Bitcoin, but BC is not only Bitcoin. Its usage goes far beyond \cite{ref03, ref04, ref05}.

In a BC network, a consensus protocol needs to be implemented to ensure any disruptive action from an adversary will be negated by a majority of participants [2]. The protocol is to decide which player among the participants in the BC network has permission to append a new block; other participants are able to verify the permission and update their local ledgers accordingly; which establishes consensus over the network \cite{ref06, ref07}. Proof of Work (PoW) is the most common consensus model used in public BCs. Unfortunately, PoW is computation-intensive, as it requires powerful nodes (i.e., miners) dedicate to solve a computationally intensive puzzle (i.e., mining), in order to produce a new block to the chain \cite{ref08}. To overcome latency and throughput bottlenecks of PoW, alternative consensus models have been proposed, including Proof of Stake (PoS) \cite{ref09, ref10}, Byzantine fault-tolerant (BFT) variants \cite{ref11}, Proof of Elapsed Time (PoET)\footnote{https://sawtooth.hyperledger.org/docs/core/releases/latest/index.html}, and Algorand \cite{ref12}. Nonetheless, such consensus protocols depend on several assumptions and impose their own disadvantages which results in limited usage in the real-world compared to the PoW-variant mechanisms \cite{ref07}.

\subsection{Smart Contracts}
An SC is a computer program deployed onto a BC network. It automatically executes "actions" when necessary "conditions" are met, specifying business logic of a service that participants have agreed to \cite{ref13}. As a mutual agreement, the content of the SC is accessible to all participants \cite{ref14}. An SC is a form of decentralised automation that facilitates, verifies, and enforces an agreement in a transaction and records the results (i.e., state changes) into a ledger. All BC frameworks have built-in mechanisms for executing SCs from a simple stack-based scripting system (e.g., Bitcoin) to a Turing-complete system (e.g., Ethereum and Hyperledger). Ethereum is among the first BCs offering Turing-completeness. Its SCs are written in either Solidity, Serpent or LLC, before being compiled to bytecodes and executed in an Ethereum Virtual Machine (EVM) \cite{ref15}. The EVM keeps track of resources consumed by the execution (i.e., $gas$) and charges to the sender's account as an incentive for miners. Hyperledger does not have its bytecode for SCs. Instead, its SCs are language-agnostic programs which are then compiled into native code, packed, installed and executed inside Docker containers \cite{ref16}. As a result, this language-agnostic design supports multiple high-level programming languages such as Go and JavaScript \cite{ref17}.

\subsection{Related Work}
Besides cryptocurrencies, the use of BC in other areas has been intensively carried out over the last few years. Specifically, prominent features of BC such as immutability, traceability, transparency, and pseudo-anonymity can be preserved for a wide range of decentralised applications (DApps), especially for managing and accounting digital assets. For instance, several projects have utilised BC in supply-chain and logistics to provide provenance tracking mechanisms for products leveraging its immutability and traceability features \cite{ref18, ref19, ref20}. The immutability and transparency features have also been utilised in a cloud data provenance platform called ProvChain \cite{ref21} in which all data operation history was transparently and permanently recorded into a BC.

Furthermore, SCs deployed in a BC framework provide autonomous functionalities executed in a decentralised manner for a wide range of domain services. Blockstack \cite{ref22} took advantage of BC for managing domain names to replace the traditional centralised Domain Name System. This work introduced pivotal functionalities including identity and discovery mechanisms deployed on top of the Namecoin platform \cite{ref23} and integrated with an off-chain storage service. In Blockstack, domain name registration and modification operations were implemented in BC whereas payload and digital signatures were stored in a Kademlia\footnote{https://en.wikipedia.org/wiki/Kademlia} Distributed Hash Table (DHT), which was connected to a virtual-chain that separated off-chain storage and BC operations. Only hashes of "name-data" tuples and state transitions were recorded on-chain. This design of decoupling the storage layer from the BC has paved the way to other studies, particularly in large-scale Internet of Things (IoT) data management \cite{ref24, ref25}. In these studies, data generated from IoT devices were stored in a DHT system and only keys of the data were recorded onto a BC. DHT nodes, responsible for managing IoT data, are required to join the BC network and listen to transactions for sending/retrieving data to/from legitimate IoT devices. BigchainDB \cite{ref26} further provided a mechanism to balance between on-chain and off-chain storage to achieve advanced features from both BC and distributed databases by using Tendermint\footnote{https://tendermint.com}, a weak synchronisation BC engine built on a BFT consensus.

Besides general-purpose data storage, BC-based accounting and management mechanisms (e.g., IdM, authorisation, access and permissions control) have also been proposed in a variety of scenarios. Lee proposed a BC-based cloud ID service for IdM \cite{ref27}, which used public-key cryptography for pseudo-identity and a distributed ledger for recording public keys. This study introduced a concept of mutual authentication by combining signatures from a client and an SP for granting access to a service. A fast security authentication scheme based on permissioned BC was proposed by Chen \textit{et al.} in a 5G ultra-dense network \cite{ref28} by using an optimised Practical BFT (PBFT) consensus protocol called APG-PBFT. APG-PBFT propagated authentication results embedded in BC among a group of access points, resulting in reducing the authentication frequency. In \cite{ref29}, a distributed access control in the IoT was proposed, with operations embedded in an SC on a public BC (i.e., Ethereum). However, most of these studies only presented high-level system design, without technical details to demonstrate the feasibility of their proposed solutions. Some platforms (e.g., \cite{ref29}) relied on a set of management nodes to play as a hub for access control, which in fact turns into the scenario of centralised management.

A few studies in the literature concerning BC-based personal data management, particularly in supporting SPs to comply with the new GDPR legislation. In \cite{ref30}, Wang \textit{et al.} proposed a fine-grained access control scheme deployed in the Ethereum framework, for personal files stored in a distributed file system called Interplanetary File System (IPFS) \cite{ref31}. It customised an attributed-based encryption scheme, but the dependency of a centralised trusted private key generator is eliminated by leveraging BC. The main limitation of this system is data owners were responsible for all required tasks, from secret key generation, file encryption, to the establishment of a secure channel for communicating with another party. The Ethereum framework was just used as a medium to execute SCs in which crypto-artifacts were embedded for identity authentication. Zyskind and Nathan \cite{ref32} proposed another access control scheme for a privacy-preserving personal data sharing platform, taking advantage of immutability and public-key cryptography in BC for identity verification and authorisation mechanisms. Similar ideas were proposed for Electronic Heath Records (EHRs) access control using Ethereum \cite{ref33, ref34} or a permissioned BC \cite{ref35}. In these works, EHRs were stored off-chain in secure data custodians whereas access control was carried out on a BC using a digital signature scheme. Neisse \textit{et al.} \cite{ref36} proposed a BC-based approach for data accountability, resulting in GDPR-compliance. They discussed different design choices respecting who create and manage data usage SCs. Similar ideas can be found in \cite{ref37, ref38}. However, in these studies, only the conceptual approach was presented; technical details on platform development were missed out. The challenges including ledger data models and functionalities in SCs have not been addressed.

%% file: 3Challenges.tex
\section{Personal Data Management: Scenarios and Challenges}
In this section, we provide an overview of the scenarios and the current solution approach on personal data management which leverages a delegated authentication and authorisation server following the OAuth standardisation \cite{ref39} (illustrated in Fig. \ref{fig1}). This solution approach is designed under a centralised client-server mindset that imposes unsolvable challenges in complying with the new GDPR requirements and in establishing trust with clients \cite{ref_add_01}.

\subsection {Scenarios}
We consider real-world scenarios in which clients allow an SP to collect, manage, process, and (possibly) shares their personal data in exchange for a service. These scenarios specify three roles as follows:

\begin{itemize}
	\item	\textit{End-user}: a client of a service who owns personal data. The end-user allows the SP to collect its data once using the service.	In the GDPR terminology, an end-user is a DS.
	\item \textit{Service Provider (SP)}: an entity that directly collects and manages personal data for its operational and business-related purposes. An SP stores personal data in an RS, which is either a system run by the SP or an independent service. An SP may share collected data with third parties for its benefits. In the context of GDPR, an SP plays both roles of a DC (when the SP shares personal data with a third-party) and a dDP (when the SP processes personal data for its own business).
	\item \textit{Third-party (TP)}: an entity that provides a service to end-users but has to rely on the SP' infrastructure to develop the service and to acquire desired personal data. In the GDPR terminology, a TP is a DP.	
\end{itemize}

\begin{figure}[!htbp]
\centering
\captionsetup{justification=centering}
	\includegraphics[width=0.45\textwidth]{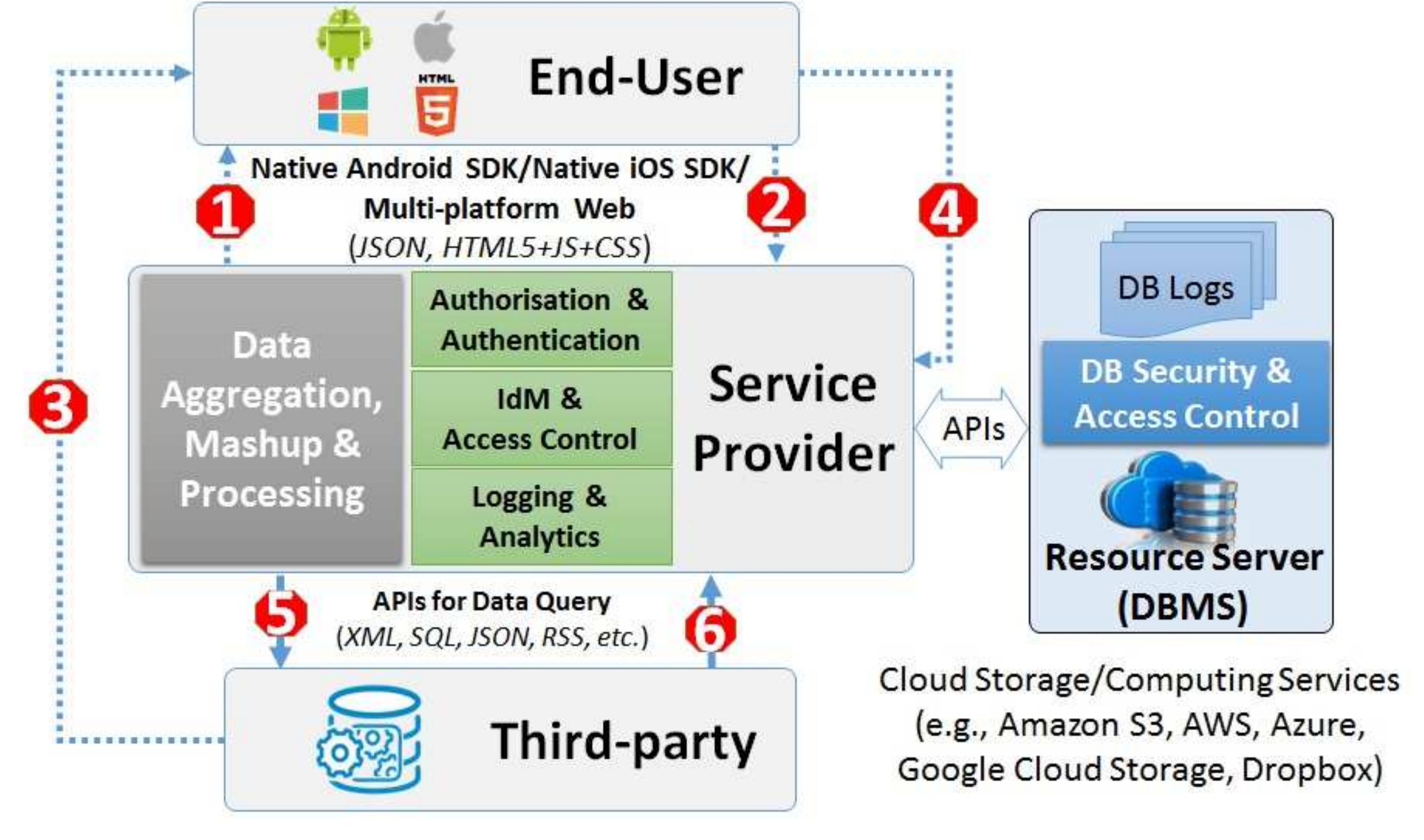}
	\caption{Personal data management and sharing scheme in the conventional client-server architecture}
	\label{fig1}
\end{figure}

As illustrated in Fig. \ref{fig1}, the procedure of granting data access for an SP and a TP is in four steps:
\begin{enumerate}
	\item A user starts to use a service provided by an SP. The SP asks the user for permission to collect her personal data.
	\item The end-user grants a set of permissions to the SP for personal data collection and processing.
	\item The TP asks the end-user to access her personal data which is collected and managed by the SP.
	\item	End-user logs into the service provided by the SP and consents a set of permissions to the TP
\end{enumerate}

Once the permission is granted, the data access procedure is in the fifth and sixth steps in Fig. \ref{fig1}:
\begin{enumerate}
	\setcounter{enumi}{4}
	\item	The SP authenticates and authorises the TP for accessing the data and provides an access token to the TP. 
	\item	The TP then calls associated APIs using the provided token in step-5 to obtain the desired data.
\end{enumerate}

\subsection {Challenges}
To meet the new GDPR requirements, conventional solutions on personal data management provide additional measures such as offering end-users mechanisms to fully control their data. Nevertheless, these measures are based on the client-server architecture which provide limited transparency and are lack of trust. For instance, a majority of SPs follow the $OAuth2$\footnote{https://oauth.net/2/} standard for access delegation, which includes IdM, authentication, authorisation, and access control mechanisms that allows end-users to share their personal data with single sign-on in a simplified and secure manner \cite{ref39}. However, the centralisation of the current approaches poses severe concern \cite{ref40}: it fully relies on the truthfulness of the SP (i.e., a delegated authentication and authorisation server) as it is the only authority to \textit{(i)} authenticate and authorise participants; and \textit{(ii)} control data access and provenance.

From an end user's perspective, this leads to a lack of transparency and accountability of data management and raise risks of personal data leakage. As all data management mechanisms are operated in a centralised system and under the SP's control, the SP may still be able to hand over personal data to an unauthorised TP without the end-user's knowledge, as far as it is not investigated by supervisory authorities. From an SP's perspective, as investigation from supervisory authority is occasionally carried out, it is challenging for an SP to declare that it has been continuously, securely and legally processing all personal data as required. This is of paramount importance for any SP to build trust with prospective clients. Furthermore, delegated permissions on personal data are not flexible as end-users do not have a fine-granular access control to impose their preferences on data usage except simple conditions predefined by SPs. Indeed, SPs currently provide only options to either "accept all" or opt-out.

Motivated by such challenges, our ultimate goal is to develop a GDPR-compliant personal data management platform by leveraging the state-of-the-art BC and SC technologies. The use of BC with SC provides autonomous operations securely executed in a decentralised manner. Furthermore, the prominent features of the BC technology, namely immutability, traceability, transparency, and pseudo-anonymity, can be effectively utilised to manage personal data fully complying with the GDPR legislation.

%% file: 4DesignConcept.tex
\section{Design Concept}
In this section, we propose a design concept for a GDPR-compliant personal data management platform, including a high-level system architecture, design guidelines, and detailed functionalities and algorithms.

\subsection{Conceptual Model and System Architecture}
\subsubsection{Assumption}
The design of a BC-based platform depends on the security models of the parties involved. In this article, we assume that an RS is "honest-but-curious" whereas SPs follow a malicious model. This means the RS executes required protocols honestly, even though it might be curious about the results it receives after the operations. If an SP correctly follows the required protocols; it will be compliant with the GDPR; otherwise violations will be logged in an immutable ledger as a record of GDPR infringements.

\subsubsection{High-level System Architecture}
\begin{figure}[!htbp]
\centering
\captionsetup{justification=centering}
	\includegraphics[width=0.45\textwidth]{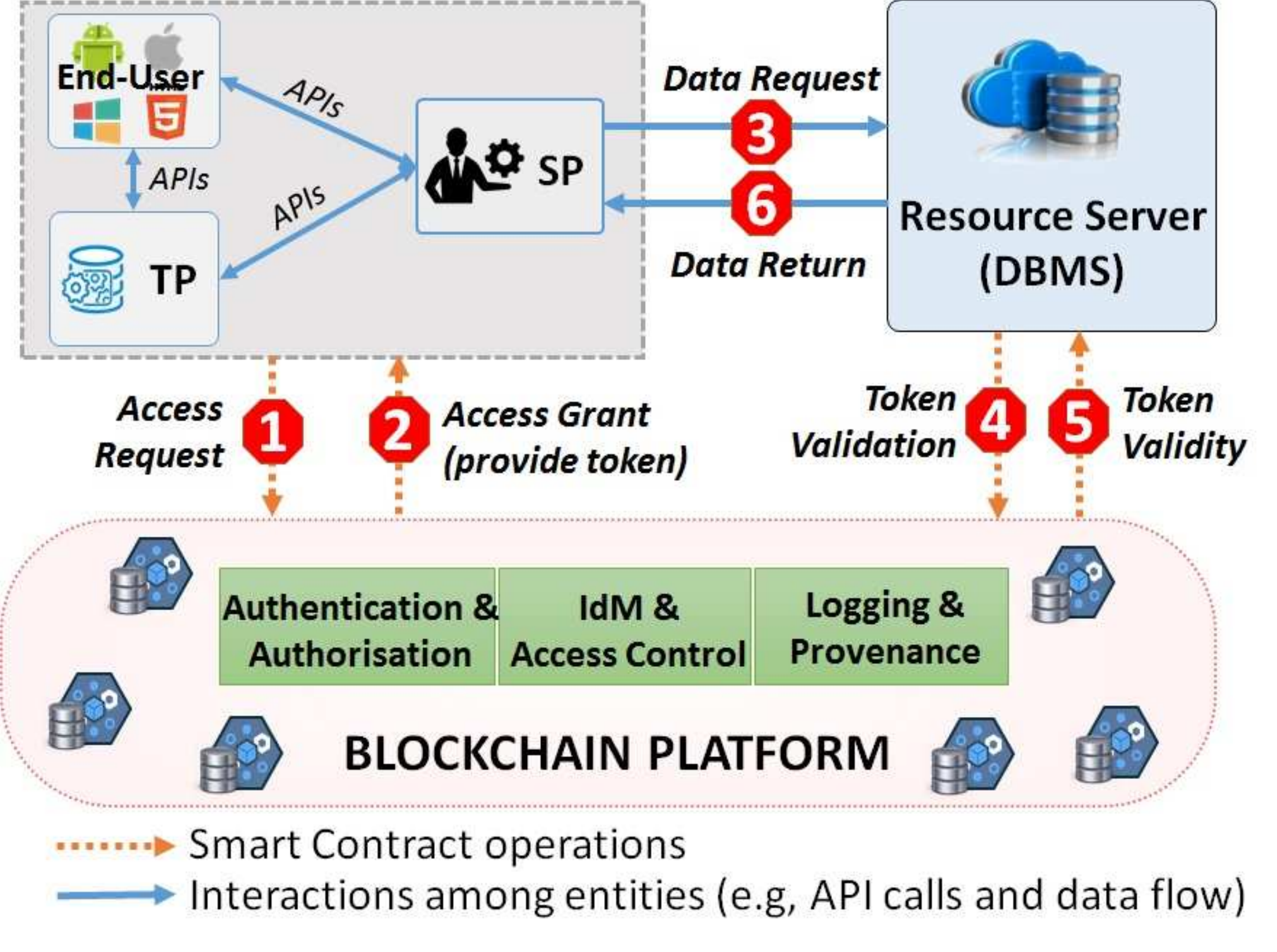}
	\caption{High-level system architecture of the design concept for a BC-based personal data management platform. The operation flow consists of 6 steps, among which step 1, 2, 4, and 5 are dedicated to granting and validating permissions operated through Smart Contracts. Step 3 and 6 operated via API calls and data-flow from/to an Resource Server.}
	\label{fig2} 
\end{figure}

A conceptual model of the proposed platform is illustrated in Fig. \ref{fig2}. The inclusive idea is that mechanisms which are related to GDPR compliance are ported to a BC network from a traditional centralised server. In particular, the Authorisation and Authentication, IdM and Access Control; and Logging and Provenance components are implemented in the form of SCs deployed in a BC network. If a BC framework offers Turing-completeness (e.g., Ethereum and Hyperledger Fabric), GDPR-related mechanisms can be conveyed by SCs. As depicted in Fig. \ref{fig2}, all activities on personal data are authenticated and authorised by the proposed BC platform (step 1 and 2). The BC, playing as a role of a delegated authentication and authorisation server, issues an \textit{access token} as "proof of permission" showing that a party has been granted to access a particular dataset. An authorised SP receives the \textit{access token} (step 2) and use it to request desired data from the RS (step 3). The RS interacts with the BC platform to validate the granted access (step 4 and 5) before returns the requested data (step 6). The validation ensures the granted access is still valid and honestly used by the corresponding authorised party.

\subsection{Design Guidelines}
\subsubsection{IdM, Authentication and Authorisation Mechanisms}
IdM, authorisation, and authentication mechanisms are of paramount importance in any data management system since they are directly related to security and privacy of the system. In the design concept, an entity in a BC network should be uniquely identified using a public-key (or hash of the public-key) in an asymmetric cryptography key-pair; authentication and authorisation processes should be implemented leveraging public-key cryptography techniques (e.g., digital signatures and encryption). In the case of permissioned BC, an additional access control layer is consolidated by using a Certificate Authority (CA) and a Membership Service Provider (MSP).

\subsubsection{Design of Distributed Ledgers}
Content of a distributed ledger reflects historical and current states of information recorded in the ledger maintained by the BC network. A personal data management platform should clarify what information and associated data model to be stored in the ledger.

\begin{enumerate}[(i)]
\item Information required to be tamper-resistant, transparent and traceable should be recorded in a distributed ledger.
    
\item[-] Any personal dataset should be specified by both DS and DC using digital signatures in a distributed ledger; 
\item[-] Data Usage Policy should be clearly specified and recorded in a distributed ledger;
\item[-] Data activities should be logged in a distributed ledger. The logs should contain information about `who', `why', `when', `what' and `how' personal data was processed;
\item[-] Hash of personal data can be recorded in a distributed ledger for data integrity checking.

\item The design of a distributed ledger must ensure:

\item[-] Designated nodes in the BC network are able to verify whether an entity is the DS or the DC of a dataset;
\item[-] Designated nodes in the BC network should be able to verify whether an entity's activity satisfies the data usage policy as recorded in a distributed ledger
\end{enumerate}

\subsubsection{Data Usage Policy}
The policy specifies data governance measures including rights, permissions, and conditions.	The usage policy should be defined in a fine-grain and expressive way using a policy language such as eXtensible Access Control Markup Language (XACML) and Model-based Security Toolkit (SecKit) designated for the IoT domains \cite{ref41}. By nature, a blockchain-based personal data management following the proposed design concept provides a fine-grained access control capability as an individual user is able to customise her own policy on each dataset by imposing access control preferences recorded onto the ledger.

\subsubsection{Off-chain Data Storage}
Personal data should be stored off-chain for better scalability and higher efficiency. Moreover, storing personal data directly onto BC, even in an encrypted form, could pose potential privacy leakage and result in non-compliance with the GDPR \cite{ref42}. Depending on specific scenarios, a conventional DBMS (e.g., Oracle or MongoDB), a storage cloud service (e.g., S3, AWS or Azure), or a distributed storage system (e.g., IPFS \cite{ref31} or Storj \cite{ref43}) can be used for data storage. Only reference to the data is stored on-chain (i.e., stored in distributed ledgers). The reference is called $data\_pointer$ that can be a hash\footnote{Hash is a type of the $data\_pointer$ used in a content-addressed storage system such as DHT, IPFS, and Stoij.}, a connection string, an absolute path, or an identifier referring to a dataset; depending on specific off-chain storage system used in the platform.

\subsection{Functionalities, Ledgers Data Model and Algorithms}

\subsubsection{Identity Management}
We introduce \textit{complex-identity}, denoted as \textit{c-ID}, to specify a digital asset associated with two or more parties. A \textit{c-ID} can be considered as an extension of asymmetric keys. In the context of the personal data management, a \textit{c-ID} of a dataset $m$ comprises an asymmetric key pair of the DS, an asymmetric key pair the DC, and an asymmetric key pair of the data pointer (denoted as $p_m$) of $m$. As the data usage policy depends on the requester's role (i.e., DS, DC, or DP), the way we define \textit{c-ID} specifies the entities associated with $m$, and simplifies the process of verification. Any digital signature scheme such as Digital Signature Algorithm (DSA) or Elliptic Curve Digital Signature Algorithm (ECDSA)\footnote{https://en.wikipedia.org/wiki/Elliptic\_Curve\_Digital\_Signature\_Algorithm} can be used to generate and manage the \textit{c-ID}, which is formally defined as a triple of probabilistic polynomial-time algorithms $(\mathcal{G},\mathcal{S},\mathcal{V})$:

\begin{itemize}
  \item $\mathcal{G}$: a key generator that creates a public-private key pair $(pk,sk)$.
	\item $\mathcal{S}$: a signing algorithm that takes $sk$ and a message $x$ as inputs and produces a signature $t=S(sk,x)$ as the output. 
	\item $\mathcal{V}$: a signature verifying algorithm that takes $pk, x, t$ as inputs, and outputs $accept$ or $reject$. For all $x$ and $(pk,sk)$, $V(pk,x,S(sk,x))=accept$.
\end{itemize}

A \textit{complete c-ID} is defined as a 6-tuple as follows:
\begin{equation}
	\label{eq_1}
	c\mhyphen ID^{comp}_{DS, DC} = (pk_{DS},sk_{DS},pk_{DC},sk_{DC},pk_{enc},sk_{enc})
\end{equation}
where $(pk_{DS},sk_{DS})$, $(pk_{DC},sk_{DC})$ and $(pk_{enc},sk_{enc})$ are asymmetric key-pairs of $DS$, $DC$ and $p_m$, respectively. The \textit{c-ID} is externally observed by nodes in a BC network as a 3-tuple:
\begin{equation}
	\label{eq_2}
	c\mhyphen ID^{ext}_{DS, DC} = (pk_{DS},pk_{DC},pk_{enc})
\end{equation}

The \textit{c-ID} is observed by the $DS$ (or $DC$) as a 5-tuple:
\begin{equation}
	\label{eq_3}
	c\mhyphen ID^{DS}_{DS, DC} = (pk_{DS},sk_{DS},pk_{DC},pk_{enc}, sk_{enc})
\end{equation}
\begin{equation}
	\label{eq_4}
	c\mhyphen ID^{DC}_{DS, DC} = (pk_{DS},pk_{DC},sk_{DC},pk_{enc}, sk_{enc})
\end{equation}

When a DS grants consent to a DP to access $m$, the private key $sk_enc$ of $p_m$ is shared to the DP through a secure channel. The DP then observes the \textit{c-ID} as a 4-tuple:
\begin{equation}
	\label{eq_5}
	c\mhyphen ID^{DP}_{DS, DC} = (pk_{DS},pk_{DC},pk_{enc},sk_{enc})
\end{equation}

The $c\mhyphen ID^{DP}_{DS, DC}$ includes the key-pair $(pk_{enc},sk_{enc})$  used to encrypt and decrypt sensitive information, including the data pointer $p_m$. Thus, only designated nodes are able to decrypt the ciphertext using the shared private key $sk_{enc}$. As a result, the information is protected from all other players in the system. Normally, RSA (Rivest-Shamir-Adleman) is used for public-key encryption mechanisms in a digital signature scheme such as DSA and ECDSA, formally defined as a 4-tuple $(\mathcal{G}, \mathcal{D}, \mathcal{E}, \mathbb{D})$: the key generator, key distribution, encryption and decryption mechanisms, respectively.

\subsubsection{Distributed Ledgers Data Model}
In the proposed design concept, ledgers are in the form of key-value pair, which is widely used in BC frameworks including Ethereum and HLF. For complex business logic, extra tasks might be required for mapping high-level data structures into key-value pairs. A state is a snapshot of a ledger at a specific time whereas state transitions are a result of transactions for creating, updating or deleting key-value pairs. A ledger contains a full history of state transitions recorded in a BC, thus it is timestamp-sequenced, immutable and tamper-resistant. With the key-value data format, all information can be obtained by referring to the latest state of the ledger, which is written in the most recent block of the BC. Some frameworks duplicate the latest state of a ledger (i.e., world-state) from a BC to a DBMS for better performance and for supporting advanced query capability (e.g., rich query). For example, either CouchDB\footnote{http://couchdb.apache.org} or LevelDB\footnote{http://leveldb.org} are used in the HLF for its world-state database.

Following the design guidelines for distributed ledgers, we specify data models for two separate ledgers used in personal data management: $3A\_ledger$ (Listing \ref{lst1}) and $log\_ledger$ (Listing \ref{lst2}). The $3A\_ledger$ is used in authentication, authorisation and access control whereas the $log\_ledger$ is used for \textit{access token} validation and logging. Both ledgers are in key-value format in which $keys$ in the $3A\_ledger$ and $log\_ledger$ are $c\mhyphen ID_{DS,DC}$ and $c\mhyphen ID_{DS,DC,DP}$, respectively. The $value$ in both ledgers contains information being used in the personal data management and provenance operations.
\begin{listing}
\begin{minted}[frame=single,
               framesep=1mm,
               linenos=true,
               xleftmargin=15pt,
               fontsize=\footnotesize,
               tabsize=10]{json}
{"3A_ledger": {
    "key": {
      "owner": pk_DS,
      "controller": pk_DC
    }
    "value" {
      "en_pointer": 3erwf3ese6d5c4...,
      "policy": {
        "rule": {Effect},{Condition},
        "action": "read, update",
        "target": "{pk_1, pk_2, ...}"
      },
      "pk_enc": "fMA0GCSqGSIb3...",
      "hash": "369f2e3e69dc40543...",
      "timestamp": 1549480378
}}}
\end{minted}
\centering
\caption{A state of the $3A\_ledger$ in JSON format. Content of the ledger includes $en\_pointer$: ciphertext of a data pointer; $pk\_enc$: public key used to encrypt the $en\_pointer$; $policy$: data usage policy, and $hash$ of the data.} 
\label{lst1}
\end{listing}

\begin{listing}
\begin{minted}[frame=single,
               framesep=1mm,
               linenos=true,
               xleftmargin=15pt,
               fontsize=\footnotesize,
               tabsize=10]{json}
{"log_ledger": {
    "key": {
      "owner": pk_DS,
      "controller": pk_DC,
      "processor": pk_DP
    }
    "value" {
      "access_token": "aAD0Gdfs234S3...",
      "issued_at": 1549480378,
      "status": "approved",
      "operation": op, 
      "scope": []ops,
      "expires_in": 3600, 
      "refresh_count": 1,
}}}
\end{minted}
\centering
\caption{A state of the $log\_ledger$ in JSON format. Content of the ledger includes $status$: either $approved$ or $rejected$, $operation$: an activity a $DP$ used to process the data such as CRUD, $scope$: a set of allowed permissions, $expires\_in$ and $refresh\_count$: dedicated to controlling the $access\_token$.} 
\label{lst2}
\end{listing}

Note that the content of the ledgers can be seen by corresponding nodes in the BC network, either honest or malicious ones. Therefore, sensitive information should be protected. For instance, asymmetric cryptography is used for pseudo-anonymous identity; and reference to a dataset (i.e., data pointer $p_m$) is encrypted (Eq. \ref{eq_6}).

\begin{equation}
	\label{eq_6}
	en\_pointer = \mathcal{E}(pk_{enc}, p_m)
\end{equation}

\subsubsection{Authentication, Authorisation and Access Control}
Public-key cryptography has been commonly used in BC-based systems to authenticate participants involved in a variety of tasks from consensus protocol participation to SC operations. In our design concept, the authentication is achieved by using the algorithm $\mathcal{V}$ in the 3-tuples digital signature scheme $\mathcal{G},\mathcal{S},\mathcal{V}$ based on any RSA/DSA-variants. The authorisation in personal data management is to specify access control (e.g., consent and usage policy); and data provenance tracking is to log data activities in an immutable and tamper-free ledger.

\begin{figure}[!htbp]
\centering
\captionsetup{justification=centering}
	\includegraphics[width=0.49\textwidth]{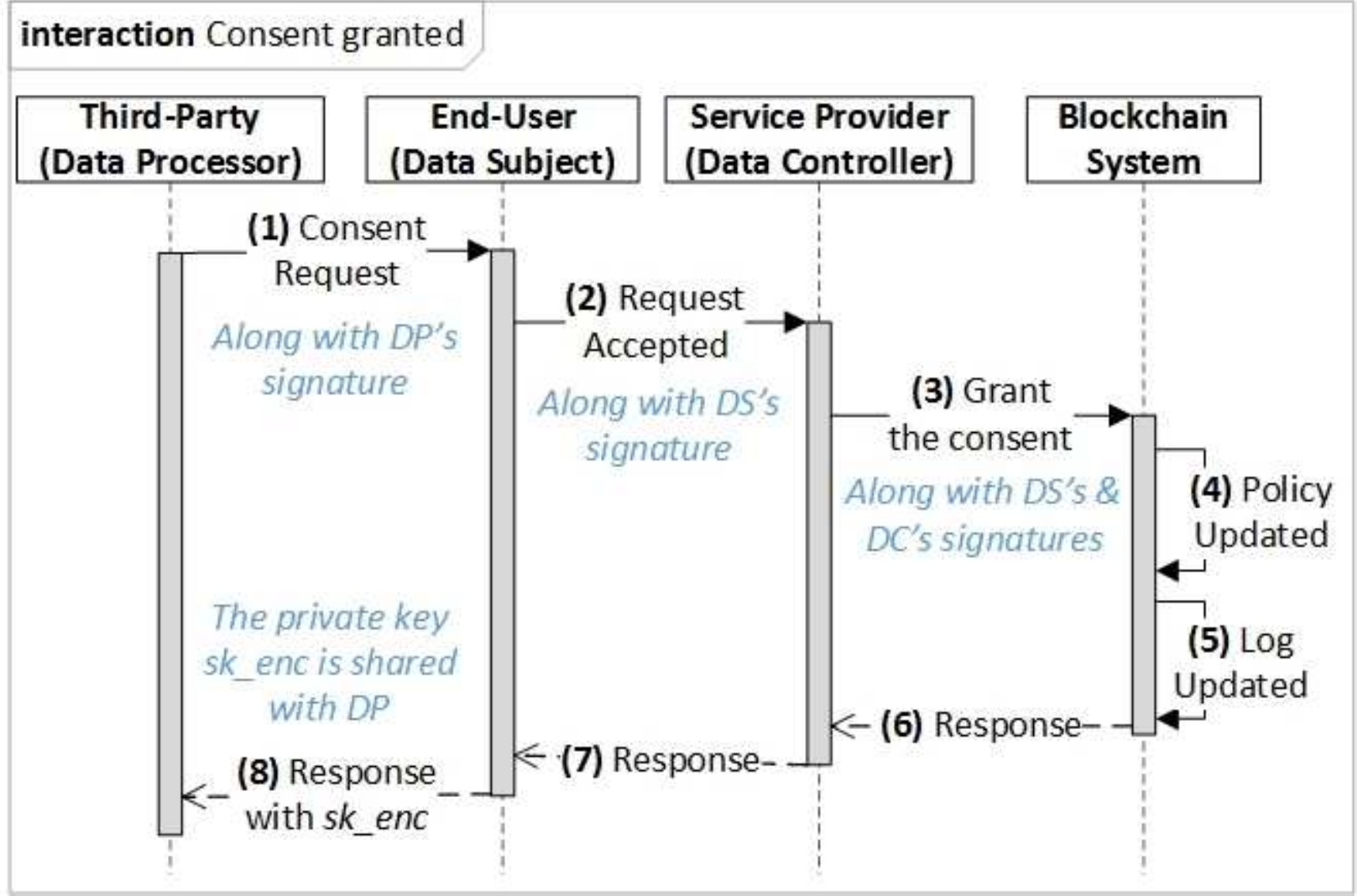}
	\caption{Process of granting consent for a DP.}
	\label{fig3}
\end{figure}

In the initial step (i.e., $Registration$ function), a $DS$ grants consent to a $DC$ for managing her personal data along with a shared key-pair $(pk_{enc},sk_{enc})$. A new record is appended into the $3A\_ledger$ specifying a new key-pair for the personal dataset with default settings granting DS all permissions (e.g, CRUD operations) specified in the $policy$. The $policy$ can be considered as an access control list/rules for a dataset, updated when consent is granted or revoked. The $hash$ and the $en\_pointer$ in the record are then updated once the DS upload her data to an RS by calling $DataUpload$ function. In our pseudo-codes, interactions with BC is through either $GetState$ or $PutState$ function provided by built-in APIs.

\begin{algorithm}
    \footnotesize
    \SetKwInOut{Input}{Input}
	\SetKwInOut{Output}{Output}
	\Input{c-ID $ci$, signature $t_{DS}$, signature $t_{DC}$, public-key $pk_{DP}$, signature $t_{DP}$, permission $op$}
    \Output{$out$}
	\textbf{Initialisation:} $rec$ $\leftarrow$ $null$, $out$ $\leftarrow$ $error$ \\
	\BlankLine
	$s1 \leftarrow \mathcal{V}(ci.pk_{DS}, t_{DS})$ \\
	$s2 \leftarrow \mathcal{V}(ci.pk_{DC}, t_{DC})$ \\
	$s3 \leftarrow \mathcal{V}(pk_{DP}, t_{DP})$ \\
	\If{($s1 \land s2 \land s3$)} {
	    $policy$ $\leftarrow$ GetState($3A\_ledger$).GetPolicy($ci$) \\
	    PutState($3A\_ledger$).Update($ci$, $policy$, \{$pk_{DP}$, $op$\}) \\
	    \BlankLine
	    $rec$ $\leftarrow$ JSON.Marshall(\{$ci, pk_{DP}$\}, \{$scope[]$+=$op$, $access\_token$=rand(), $issue\_at$=Time.now(), $status$="approved"\}); \\
        PutState($log\_ledger$).Append($rec$); \\
        $out$ $\leftarrow$ $success$
	}
	\textbf{Return} $out$
	\caption{$GrantConsent$ grants a consent for a DP}\label{alg1}
\end{algorithm}

Fig. \ref{fig3} depicts a sequence diagram of granting consent for a DP. The consent is granted if both DS and DP accept the request by providing their digital signatures $t\_DS$ and $t\_DC$ in step (2) and (3). Step (4) and (5) are carried out by the $GrantConsent$ function (Alg. \ref{alg1}). Authentication is achieved by using verification function $\mathcal{V}$ for all DS, DC, and DP (line 2-4). If the authentication is accepted (line 5), access control is then carried out by reflecting the permission into $policy$ in the $3A\_ledger$. As depicted in Alg. \ref{alg1}, the $GrantConsent$ firstly grants permissions (i.e., requested operation $op$) by updating policy with $op$ in the $3A\_ledger$ (line 6, 7). Secondly, the $GrantConsent$ appends a new record into the $log\_ledger$ (line 9), which is used for validating and logging whenever the DP accesses the data. The $access\_token$ with other metadata is generated as $value$ in the \textit{key-value} format record (line 8). Technically, $access\_token$ is a string of random-looking characters referring to a collection of metadata in the $log\_ledger$. A multi-signature technique is also used in the algorithm to ensure consent is granted by both DS and DC.

$RevokeConsent$ function is to revoke a permission previously granted to a DP. As depicted in Alg. \ref{alg2}, it is only executed by either DS or DC. Similar to $GrantConsent$ function, $RevokeConsent$ appends an updated policy excluded the revoked permission $op$ to the $3A\_ledger$ (line 4, 5) and updates the $log\_ledger$ accordingly (line 6,7).

\begin{algorithm}
    \footnotesize
    \SetKwInOut{Input}{Input}
	\SetKwInOut{Output}{Output}
	\Input{c-ID $ci$, signature $t$, public-key $pk_{DP}$, permission $op$}
    \Output{$out$}
	\textbf{Initialisation:} $rec$ = $null$, $out$ = $error$ \\
	\BlankLine
	$s \leftarrow (\mathcal{V}(ci.pk_{DS}, t) \lor \mathcal{V}(ci.pk_{DC}, t))$ \\
	\If{$s$} {
	    $policy$ $\leftarrow$ GetState($3A\_ledger$).GetPolicy($ci$) \\
	    PutState($3A\_ledger$).Update($ci$, $policy$, \{$pk_{DP}$, $-op$\}) \\
	    \BlankLine
	    $rec$ $\leftarrow$ GetState($log\_ledger$).GetRecord($ci$, $pk_{DP}$) \\
	    $rec$ $\leftarrow$ PutState($log\_ledger$).Update($rec$, \{$scope[]$-=$op$, $access\_token$=rand(), $issue\_at$=Time.now()\}); \\
        $out$ $\leftarrow$ $success$
	}
	\textbf{Return} $out$
	\caption{$RevokeConsent$ revokes a permission previously granted to a DP}\label{alg2}
\end{algorithm}

\begin{figure}[!htbp]
\centering
\captionsetup{justification=centering}
	\includegraphics[width=0.49\textwidth]{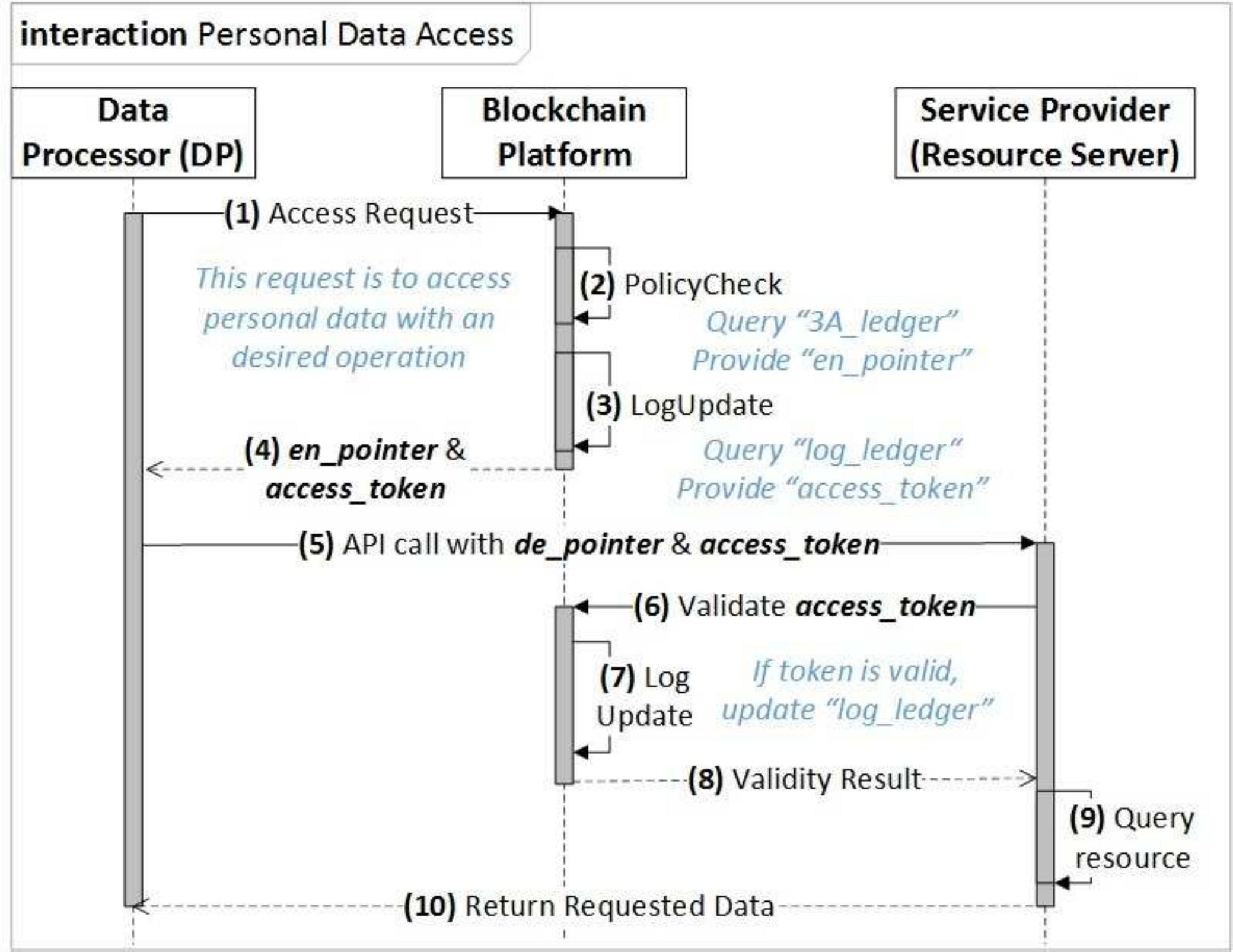}
	\caption{Sequence diagram of accessing data stored in an RS by a DP}
	\label{fig4}
\end{figure}

Once consent is grant, the operation flow of accessing personal data is demonstrated in Fig. \ref{fig4}. Whenever DP desires to access personal data (step (1)), it invokes a corresponding SC with the $DataAccess$ function (Alg. \ref{alg3}). As can be seen in Fig. \ref{fig4}, after checking eligibility of the call (i.e., step (2) and (3) executed by line 2, 3 in Alg. \ref{alg3}), the SC returns two outputs $en\_pointer$ and $access\_token$ to the DP (step (4)), executed by line 6-9 in Alg. \ref{alg3}. The DP then uses the shared private key $sk_{enc}$ (already obtained from step (8) in Fig. \ref{fig3}) for decrypting the $en\_pointer$. The decrypted ciphertext (i.e., $de\_pointer$) is the $data pointer$ for the desired dataset. Both $de\_pointer$ and $access\_token$ are used as parameters for an API call to process the data (step (5)).

\begin{algorithm}
    \footnotesize
    \SetKwInOut{Input}{Input}
	\SetKwInOut{Output}{Output}
	\Input{c-ID $ci$, public-key $pk_{DP}$, signature $t_DP$, permission $op$}
    \Output{$out$}
	\textbf{Initialisation:} $rec$ $\leftarrow$ $null$, $out$ $\leftarrow$ $rejected$ \\
	\BlankLine
	$s \leftarrow (\mathcal{V}(pk_{DP}, t_{DP})$ \\
	\If{$s$} {
	    $policy$ $\leftarrow$ GetState($3A\_ledger$).GetPolicy($ci$) \\
	    \If{($policy \subset (pk_{DP}, op)$)} {
	        $en\_pointer$ $\leftarrow$ GetState($3A\_ledger$).GetPointer($ci$);\\
	        $access\_token$ $\leftarrow$ GetState($log\_ledger$).GetToken($ci$, $pk_{DP}$);\\
	        \BlankLine
	        $out$ $\leftarrow$ ($en\_pointer, access\_token$)
	    }
	}
	\textbf{Return} $out$
	\caption{$DataAccess$ returns $en\_pointer$ and $access\_token$ for an eligible request}\label{alg3}
\end{algorithm}

A function called $TokenValidation$ is dedicated to double-checking the validity of the $access\_token$ and updates the $log\_ledger$. In Alg. \ref{alg4}, line 4 is to obtain metadata associated with the $access\_token$ from the $log\_ledger$; if the request is from DS or DC then there is no need to validate the $access\_token$; only $log\_ledger$ is updated (line 5-7). Otherwise, the validation is then conducted by inspecting the metadata (line 9-12) before updating the $log\_ledger$ (line 13). The $TokenValidation$ is performed to ensure that only API calls with valid an $access\_token$ leads to an execution of the call (step (9)). Step (7) safeguards that all valid API calls are autonomously logged in the $log\_ledger$. It is worth to mention that the honest-but-curious RS assumption plays a key role in the success of our platform because the RS must follow the authorisation process (i.e., double-check API calls from DPs with the BC system) before executing the calls.

\begin{algorithm}
    \footnotesize
    \SetKwInOut{Input}{Input}
	\SetKwInOut{Output}{Output}
	\Input{Token $access\_token$, public-key $pk$, signature $t$ permission $op$}
    \Output{$out$}
	\textbf{Initialisation:} $rec$ $\leftarrow$ $null$, $out$ $\leftarrow$ $rejected$ \\
	\BlankLine
	$s \leftarrow (\mathcal{V}(pk, t)$ \\
	\If{$s$} {
	    $rec$ $\leftarrow$ GetState($log\_ledger$).Query($access\_token$) \\
	    \If{(($rec.owner = pk$) $\lor$ ($rec.controller = pk$))} {
	        $rec$ $\leftarrow$ PutState($log\_ledger$).Update($rec$, \{$expires\_in$-=Time.now(), $issue\_at$=Time.now()\}); \\
	        $out$ $\leftarrow$ $accepted$
	    } \Else{
	        \If{(
	            ($rec.processor = pk$) $\land$ ($rec.scope \subset op$) $\land$ \\
                ($rec.expires\_in > 0$) $\land$ ($rec.operation = op$) $\land$ \\
                ($rec.status = approved$) $\land$ ...)}
            {
                $rec$ $\leftarrow$ PutState($log\_ledger$).Update($rec$, \{$expires\_in$-=Time.now(), $issue\_at$=Time.now()\}); \\
                $out$ $\leftarrow$ $accepted$\\
            }
        }
	}
	\textbf{Return} $out$
	\caption{$TokenValidation$ double-checks the validity of an $access\_token$ and update the $log\_ledger$}\label{alg4}
\end{algorithm}

%% file: 5Deployment.tex
\section{Platform Deployment in Permission Blockchain}
In this section, we implement a platform following the proposed design concept for managing personal profiles for an SNS. The choice of using a permissioned BC framework in the demonstration does not imply that a public one is less appropriate for implementing the proposed design concept. Instead, HLF is chosen due to its business-oriented architecture offering better adaptation to the use-case; also, thanks to its readily existing software components for a rapid development cycle of our platform. Detailed technical solutions and implementation of the platform are presented. Source-code of the demonstration can be obtained from Github\footnote{https://github.com/nguyentb/Personal-data-management}.

\subsection{HLF Platform Setup}
HLF is the most popular permissioned BC framework used by big enterprises such as IBM and Microsoft. As being permissioned, a node involved in an HLF network is associated with an identity and permissions provided by a CA and an MSP, respectively. Nodes in HLF take up one of three roles: \textit{Client}, \textit{Peer} and \textit{Ordering Service Nodes} (OSNs). In our demonstration and for the performance evaluation, we have deployed different HLF network settings include $3$ OSNs running in $Kafka$ cluster mode for providing the ordering service, from $4$ to $32$ peers, and a varied number of clients from $10$ to $1000$. All peer nodes endorse both SCs (i.e., chaincodes in HLF terminology), namely $3A\_cc$ and $log\_cc$. That means these two SCs are locally installed, instantiated and executed in all 5 peers to interact with the two ledgers $3A\_ledger$ and $log\_ledger$, respectively. These two ledgers are exactly following the data models described in Section IV.D. As the two distributed ledgers are being used and HLF allows only one ledger per channel\footnote{Channel is a terminology in HLF technically referring to a private blockchain overlays which offers data isolation and transaction confidentiality.}, two HLF channels are created, namely $3A\_channel$ and $log\_channel$. All Peers and OSNs belong to both channels; the $3A\_cc$ and the $log\_cc$ SCs are operated in the $3A\_channel$ and the $log\_channel$, respectively. As a result, all the peer nodes separately endorse the two SCs corresponding to different local ledgers. The two local ledgers are stored in Linux filesystem whereas the world-state database is duplicated in CouchDB.

\begin{figure}[!htbp]
\centering
\captionsetup{justification=centering}
	\includegraphics[width=0.45\textwidth]{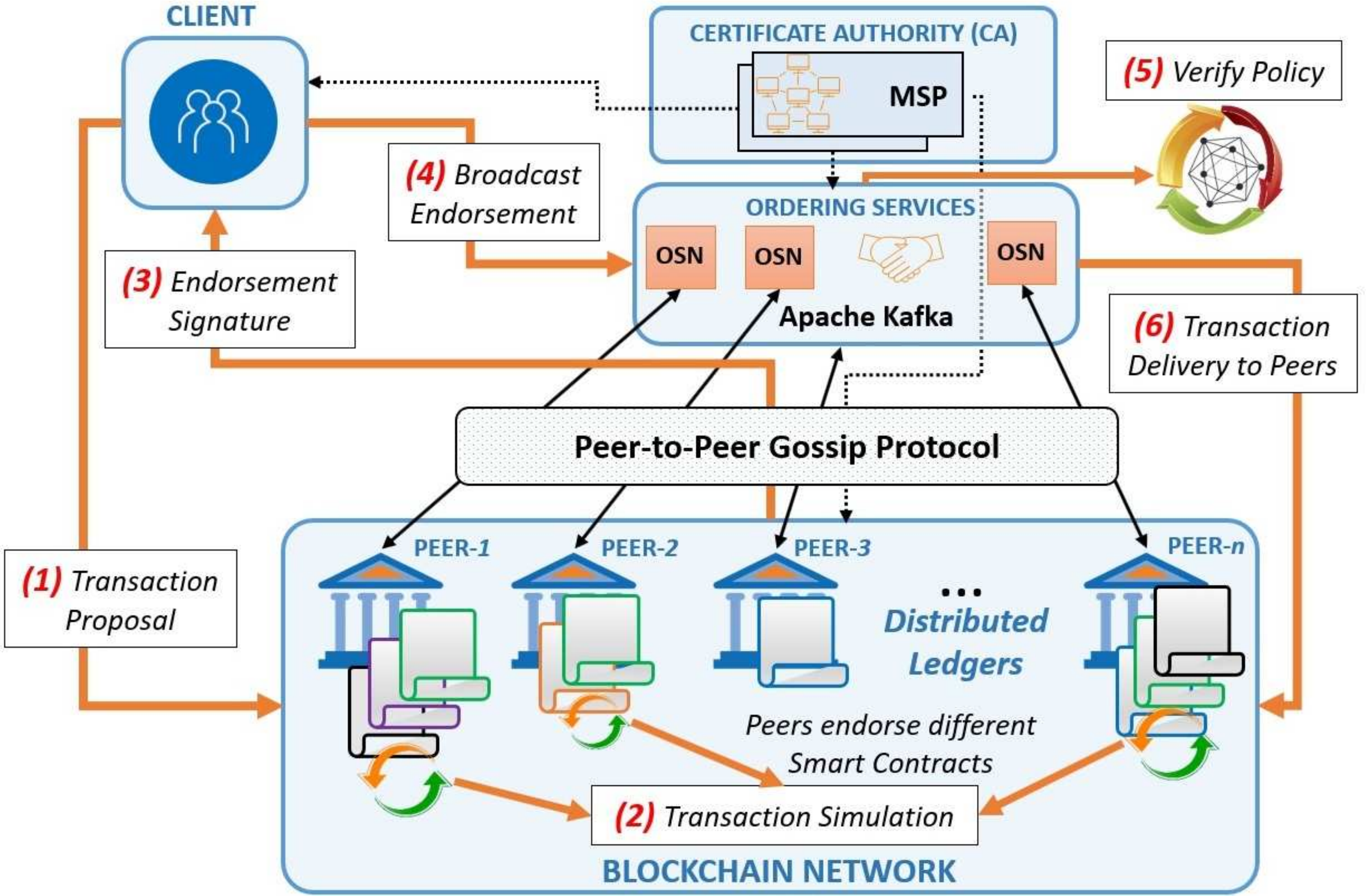}
	\caption{High-level system architecture and transaction flow of the HLF framework}
	\label{fig5}
\end{figure}

All clients are populated using the Fabric Client SDK (for NodeJS) for interacting with the HLF network. As illustrated in Fig. \ref{fig5}, a client constructs a transaction proposal to invoke either $3A\_cc$ or $log\_cc$ SCs (step-1) and sends to all endorsing peers (i.e., endorsers). These peers verify the proposal and locally execute the $3A\_cc$ or $log\_cc$ to produce an endorsement signature (i.e., transaction results with the peer's signature) (step-2) and pass back to the client (step-3). Once receiving endorsement signatures, the client assembles the endorsements into the transaction and broadcast it to the OSNs, running $Kafka$ mode (step-4). The OSNs validate and commit the transaction (step-5), then broadcast a message to all peers to update their local ledgers (step-6). In case the transaction is not successful, and the ledgers are not updated but the proposal is still logged for audit.

\subsection{Personal Profile Management Use-case}
\begin{figure}[!htbp]
\centering
\captionsetup{justification=centering}
	\includegraphics[width=0.45\textwidth]{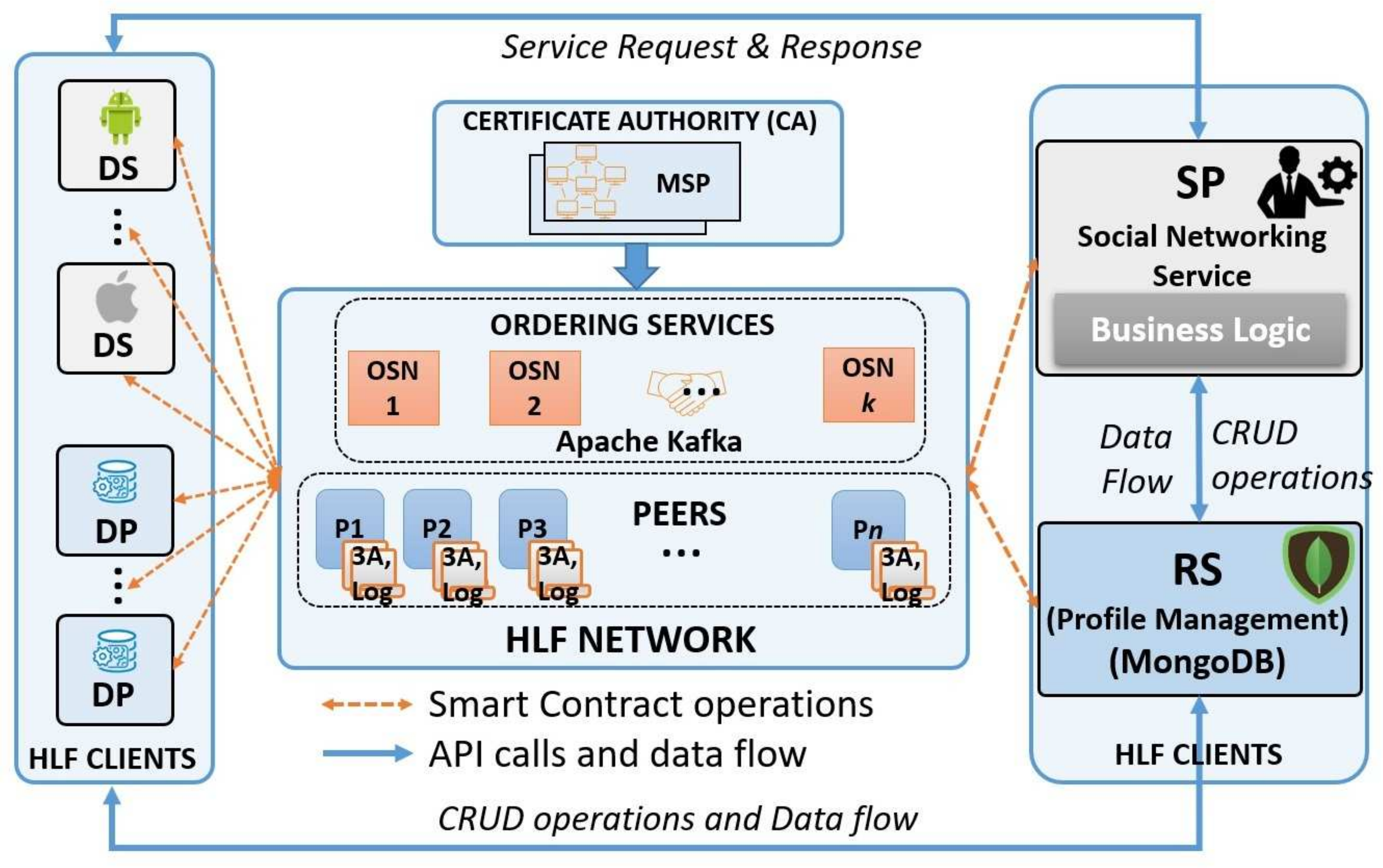}
	\caption{System architecture of a GDPR-compliant social networking service with the RS for personal profiles using HLF}
	\label{fig6}
\end{figure}

We consider a use-case that a social networking SP processing profile data stored in a separate RS. This RS follows the honest-but-curious model anticipating the BC as an HLF client and honestly executing required protocols (i.e., interacting with the BC network for token validation). To comply with the GDPR, the SP participates in the proposed BC-based platform (Fig. \ref{fig6}). To demonstrate the use-case, we build the RS as a profile management web-service based on REST architecture\footnote{https://en.wikipedia.org/wiki/Representational\_state\_transfer} for parties to process profile data through calling corresponding RESTful APIs. Profile information is stored in JSON-like documents using MongoDB\footnote{https://www.mongodb.com/}, a document-oriented database system. The profile data model follows the Friend-Of-a-Friend (FOAF) ontology for describing person which is normally used in social networks\footnote{http://xmlns.com/foaf/spec/}. Processing a profile includes $\{create, read, update, delete\}$ CRUD operations by making a request to a corresponding API provided by the RS.

A request to a RESTful API contains 6 parameters: $(1)API\mhyphen Endpoint$, $(2)REST\mhyphen Endpoint$, $(3)Method$, $(4)Header$, $(5)Params$, $(6)Payload$ in which the first fours are required. A RESTful request is as follows:

\begin{listing}
\begin{minted}[frame=single,
               framesep=1mm,
               linenos=true,
               xleftmargin=15pt,
               fontsize=\footnotesize,
               tabsize=10]{json}
POST localhost:8080/ProfileManagement 
-H 'Content-Type:application/json' 
pubkey=pk&
signature=t&
token=access_token
&operation=read
\end{minted}
\centering
\end{listing}
where $Method$ is $POST$, $REST\mhyphen Endpoint$ is $localhost$:$8080$, $API\mhyphen Endpoint$ is $/ProfileManagement$, Header is $Content\mhyphen Type$:$application/json$ following by $Params$ including the public-key $pk$ with the signature $t$, the $access\_token$, and the requested $Read$ operation.

\subsection{Identity Management and Pseudo-anonymity}
Any entity in HLF including clients, peers, orderers, CAs and MSPs needs to be identified by digital identities (e.g., $X.509$ standard) before interacting with the HLF network. In our HLF-based system, a built-in CA called \textit{Fabric CA} is used to generate $X.509$ digital certificates, adopting the traditional Public Key Infrastructure (PKI) hierarchical model. An $X.509$ digital certificate contains a public key (along with a corresponding private key) and associated information of an entity (e.g., organisation, host-name, and domain. This certificate is then either signed by the Fabric CA or self-signed. The Fabric CA server in our system is initialised using Docker which hosts an HTTP server on the default port $7054$ that offers REST APIs. All entities have to enrol and register with the CA server via the REST APIs using either the \textit{Fabric CA client} or the \textit{Fabric SDK} before participating in the blockchain system. Once an entity is enrolled and registered, an enrolment certificate ($eCert$), a network transaction certificate ($tCert$), a CA certificate, and a corresponding private key are stored in $PEM$ files in the subdirectories of the entity's directory.

In the HLF settings, ECDSA, an updated version of the DSA scheme leveraging elliptic-curve cryptography, is used with 256-bit key-size, which guarantees that any public-private key pairs generated by the generator $\mathcal{G}$ is \textit{practically} unique across the HLF network. Moreover, the hiding property of the ECDSA also ensures that there is no practical mechanism to recover a private key from the corresponding public key \cite{ref_add_03}. As a result, HLF entities, whose identifiers are $X.509$ digital certificates, preserve the pseudo-anonymity property. However, as HLF is a permissioned blockchain, all of HLF entities are under control of a certificate authority CA (in our system is the built-in Fabric CA); this means the pseudo-anonymity property depends on the security and trustworthiness of the Fabric CA.

To administer entities evolving in variety of HLF tasks, MSP is used for specifying participants, roles, and access privileges in a HLF network and channel. An MSP provides a configuration identifying trusted root and intermediate CAs; these CAs then define members of a trust domain by either \textit{(i)} listing identities of the members or \textit{(ii)} identifying authorised CAs that issue valid identities for members. The latter is used in the demonstration. Technically, an entity's identity is associated with its MSP and implemented using the HLF client identity chaincode library $cid$\footnote{https://github.com/hyperledger/fabric/blob/release-1.1/core/chaincode/lib/cid/README.md} as shown in Listing \ref{lst3}:
\begin{listing}
\begin{minted}[linenos=true,
               xleftmargin=15pt,
               fontsize=\footnotesize,
               tabsize=10]{js}
function ClientID(stub shim.ccAPI) ci *clientID {
    hlfId = ci.New(stub);
    mspID = hlf.GetMSPID();
    cert = hlf.GetX509Certificate();
    return &clientID{mspID, cert};
}
\end{minted}
\centering
\caption{Identity of a HLF client constituted from $mspID$ and $X.509$ certificates utilising the $cid$ library}
\label{lst3}
\end{listing}

\subsection{Smart Contracts Implementation}
There are two chaincodes implemented in the HLF network: \textit{(i)} the $3A\_cc$ for authentication, authorisation and access control, operating with the $3A\_ledger$; and \textit{(ii)} the $log\_cc$ for access validation and logging, operating with the $log\_ledger$. Theoretically, a contract can be written in any programming language; and in the demonstration, $Go$ language is used. The two chaincodes inherit the built-in $shim$ package\footnote{https://godoc.org/github.com/hyperledger/fabric/core/chaincode/shim}, which provides a variety of APIs to interact with distributed ledgers such as accessing state variables, transaction context and call other chaincodes.

Regarding the distributed ledgers, $en\_pointer$ is the ciphertext of an identifier of a data object (i.e., $profile.ID$) using the encryption function $\mathcal{E}$ with the encryption key $pk_{enc}$:
\begin{equation}
	\label{eq_8}
	en\_pointer = \mathcal{E}(pk_{enc}, profile.ID)
\end{equation}

A party who permitted access a profile has a shared private key $sk\_enc$ to decrypt $en\_pointer$ in order to obtain the $profile.ID$, which is then passed as a parameter for a RESTful API to access the desired profile information:
\begin{equation}
	\label{eq_9}
	profile.ID = \mathbb{D}(sk_{enc}, en\_pointer)
\end{equation}

The $policy$ in the $3A\_ledger$ is simply defined as an access control list (ACL) as shown in Listing \ref{lst4}. The ACL is implemented as a struct in the $3A_cc$ specifying four \textit{access rights} for participants called $Create$, $Read$, $Update$ and $Delete$ (representing four CRUD operations). Associated with each \textit{access right} is a list of granted parties including DS, DC and DPs (under their public keys such as $pk\_DS$, $pk\_DC$, and $pk\_DP$).
\begin{listing}
\begin{minted}[frame=single,
               framesep=1mm,
               linenos=true,
               xleftmargin=15pt,
               fontsize=\footnotesize,
               tabsize=10]{json}
"policy" {
    "Create": {pk_DS, pk_DC, ..},
    "Read": {pk_DS, pk_DC, pk_DP1, pk_DP2, ..},
    "Update": {pk_DS, pk_DC, pk_DP3 ...},
    "Delete": {pk_DS, pk_DC, pk_DP3, pk_DP4, ..}
}
\end{minted}
\centering
\caption{Data Usage Policy defined as an Access Control List under JSON format} 
\label{lst4}
\end{listing}

Based on the identity scheme and detailed information for the two ledgers, core functions in personal data management such as $GrantConsent$, $RevokeConsent$, $TokenValidation$ and $DataAccess$ are then implemented exactly following the algorithms described in Section III.D.

%% file: 6Analysis.tex
\section{Analysis and Discussion}
This section provides analysis and discussion on the platform deployed in Section V, including GDPR-compliance applicability, threat models and system performance.

\subsection{Trust Assumption}
Besides the \textit{honest-but-curious} characteristic of the RS, a must assumption is that a large portion of peer nodes in the HLF network are honest. Generally, HLF v1.x offers multiple ordering techniques including a variety of BFT-based solutions such as pBFT. Such BFT-variant protocols are able to conditionally tolerate $\lfloor\frac{N-1}{5}\rfloor$ (e.g., in Ripple \cite{ref44}) to $\lfloor\frac{N-1}{2}\rfloor$ (e.g., in crash-fault tolerance) simultaneously faulty nodes. Such BFT-variant protocols guarantee consistency despite any number of node failures and network partition, with at most $\lfloor\frac{N-1}{3}\rfloor$ faulty nodes \cite{ref45}. Unfortunately, these protocols are under development for the HLF framework, and only Apache Kafka is provided as a reference implementation, which supports some levels of fault-tolerant (e.g., crash-faulty) but not Byzantine failure.

The cryptographic primitives (i.e., cryptographic hash function $SHA256$, the public-key cryptography RSA and the digital signature schemes ECDSA) are practically secure. This means adversaries are not able to: \textit{(i)} reverse/break the cryptographic hash function, \textit{(ii)} reverse a public key to obtain a private key, and \textit{(iii)} forge a digital signature of another party without knowing the corresponding private key. As our system is built on top of the permissioned blockchain HLF, the Fabric CA with built-in PKI, which are responsible for the distribution of management of $X.509$ digital certificates, are assumed to be secure and honest. This means in general adversaries are not able to mislead the Fabric CA in large-scale (e.g., more than $\lfloor\frac{N-1}{3}\rfloor$ adversaries granted in the HLF network) in order to subvert the HLF system (e.g., 51\% attack or Sybil attack). However, some internal adversaries might be granted to participate in the network, resulting in non-GDPR-compliance. Regarding key management, we assume that private keys obtained by the key generator $\mathcal{G}$ are effectively protected from adversaries by leveraging existing solutions from enterprise systems. However, this is the weak assumption, meaning that an adversary could somehow obtain a private key and impersonate an honest party to access data which also leads to non-GDPR-compliance. These threats of non-GDPR-compliance will be considered under Section VI.C.

\subsection{GDPR-Compliance}
From an applicability perspective, the proposed platform provides SPs (e.g., the SNS) mechanisms to fully comply with the GDPR. This is due to the following reasons:

\subsubsection{Full Control back to Data Owners}
As following the design concept, the platform provides DSs:
\begin{itemize}
\item "Right of access" and "right of rectification": This is because DS is eligible to do all CRUD operations to her personal data as specified in the default policy when ledgers are initialised, and no one can change these rights.
\item "Right of restricted processing" and "right of data portability": This is because DSs have full permissions to manage data usage policy (e.g., to grant or revoke consent anytime/anywhere by invoking the $GrantConsent$ and $RevokeConsent$ functions in the $3A\_cc$).
\item "Right to be informed": This is because the platform always requires DS's signature for data collection or for granting consent.
\item "Right to be forgotten": As personal data is stored off-chain, an RS is able to erase the data as requested from DS. However, a question is posed when leveraging BC for personal data management: "whether a BC platform complies with the GDPR as distributed ledgers are immutable; meaning that the ledgers, theoretically, will never be erased?". Therefore, if a piece of personal information is recorded in a ledger, the platform will violate the "right of forgotten". In the design concept, sensitive information is encrypted before writing into a ledger (e.g., $data\_pointer$). The "right of forgotten" is then ensured by throwing decryption keys. Whether this remedy fully satisfies the GDPR is still an open question \cite{ref42, ref46}.
\end{itemize}

\subsubsection{Security, Transparency and Accountability}
By following the design concept, the platform ensures that:
\begin{itemize}
\item Security of the identity, authentication and authorisation mechanisms, which depends on the security of the cryptographic primitives, is assumed to be secure.
\item Operations (e.g., grant or revoke a consent, update usage policy, verify \textit{access token}, and CRUD) are authenticated, authorised and autonomously executed only by invoking corresponding SCs deployed in the HLF network. This ensures system procedures are executed in a transparent and not compromised by any individuals.
\item Information about management operations and CRUD activities on personal data, including who/what/when/why/ and how are immutably recorded in the $log\_ledger$.
\end{itemize}

Consequently, the proposed platform forces SPs, who participate in the system, to be responsible for complying with the GDPR; otherwise any unauthorised or malicious transactions initiated by a corresponding SP can be always figured out. Furthermore, the investigation for GDPR-compliance is empowered as all activities logged in the ledgers can be traced back. The signalling of a non-compliant activity could trigger official investigation and auditing of an SP by a supervisory authority. The decisions could be made based on whether a malicious activity recorded in the $log\_ledger$ exists that respects the associated data usage policies in the $3A\_ledger$. In this regard, the two distributed ledgers can be considered as legal grounds for the GDPR compliance. As a result, the platform is able to demonstrate the GDPR compliance. Therefore, the proposed BC-based platform provides efficient measures to meet the requirements of data accountability. For those reasons, a social networking SP, which utilises the platform for its personal data management tasks, fully complies with the GDPR.

\subsection{Threat Models}
The advanced capability of the BC framework plays a key role in providing a secure and trustworthy platform for complying with the GDPR. However, certain aspects of the contemporary BC and SC technologies present limitations imposing threats resulting in non-compliance with the GDPR.

\subsubsection{Security Threats}
Given the aforementioned assumptions, the decentralised nature of the BC ensures that an adversary cannot corrupt the BC network to unauthorisedly change the ledgers as that would imply the majority of the network's resources are compromised. Also, the adversary cannot impersonate an authorised party as its digital signature cannot be forged. Security threats are, thus, from two sources: \textit{(i)} an internal adversary acting in a Byzantine way, who has been granted to access personal data; and \textit{(ii)} an honest party whom both private key and decryption key $sk_{enc}$ are disclosed to an external adversary; thus, the adversary could pose itself as the party. In such scenarios, the $TokenValidation$ function is of paramount importance since it plays as a role of a gatekeeper to reassure that any $access\_token$ expires after the amount of time and needs to be refreshed (i.e., re-authenticated and re-authorised). As a result, the $TokenValidation$ mitigates the risk of a long-lived $access\_token$ leaking, similar to the use of both $access\_token$ and $refresh\_token$ used in the standard OAuth2 specification\footnote{https://tools.ietf.org/html/rfc6749}.

Admittedly, it is inevitable that an adversary is able to access the data in the time-frame window of the $access\_token$ (defined by the $expires\_in$ parameter in the $log\_ledger$). During this period, it is unachievable to prevent the adversary from accessing data unless the security breach is detected. Once being detected, DS is able to revoke the consent by updating the ledgers to remove all permissions related to the adversary. The remedy is straightforward in case of the first scenario - the party is malicious. However, it turns to a complex situation when an honest party leaks its private key to the adversary. This party is never able to get granted again as its identity is compromised, which is unreasonable. A key management with an account recovery scheme could be an applicable solution to deal with this situation although it is expected to be much complicated to integrate the recovery scheme with a BC system \cite{ref47}. Another security threat comes from poor quality code in SCs which exposes vulnerabilities to be exploited. For example, an attacker stole $3.6M$ Ether (worth \$50M at that time) in DAO\footnote{https://ethereum.org/dao} attack exploiting a concurrency bug in DAO's SCs. As a BC framework supporting Turing-complete SCs, software bugs are painful to avoid. Thus, SCs must be written in high-quality standards and follows strict security specifications \cite{ref17, ref48}.

\subsubsection{Privacy Threats}
The openness of distributed ledgers, which allows parties to inspect, violates the idea of privacy. Even in a permissioned BC in which transactions take place between authenticated parties, some privacy threats remain as any participants could be malicious. In the proposed design concept, measures to tackle privacy leakage are to both: \textit{(i)} provide pseudo-anonymity for parties using public key cryptography as identities; and \textit{(ii)} encrypt sensitive information exposed on the ledgers.

The first measure provides pseudo-anonymity, thus, there is a possibility to link between public addresses with physical identification of the users by using a variety of de-anonymisation techniques \cite{ref49}. Literally, the risk of revealing real-world identity by an adversary can be significantly reduced in a permissioned BC compared to a public one thanks to an additional permission access control layer \cite{ref14, ref50}. As a trade-off, anonymity is sacrificed as it requires more identity materials for stringent privacy requirements.

The second measure is to encrypt $data\_pointer$ (i.e., $profile.ID$ in the demonstration), which is used as a parameter in API calls for accessing a personal dataset. The encryption ensures that the information is only visible to designated parties, reducing the risk of leaking the information to adversaries. Some other information recorded onto the ledgers such as data usage policy (i.e., $policy$) and activities log must be in plain text as the information is referred by peers for some business logic when executing the chaincodes. Even though this information is not directly related to identifiable individuals, this might be a source of privacy leakage as it might be used in de-anonymisation techniques. At this moment, no particular privacy threat has been pointed out due to exposing such information. Nevertheless, further investigation might need to be carried out for this potential threat. Homomorphic Encryption could be used for information encryption supporting query on cipher-text \cite{ref_add_04}. Flexible encryption schemes like attribute-based encryption (ABE) \cite{ref_add_05} might also be used as a remedy to encrypt such information and only a designated group of peers can be decrypted. These schemes are only suitable for permissioned BC as they rely on a trusted key generator - which could be integrated in a CA.

\subsection{Performance Evaluation}
\begin{figure*}[ht]
	\centering
	\captionsetup{justification=centering}
	\includegraphics[width=\textwidth]{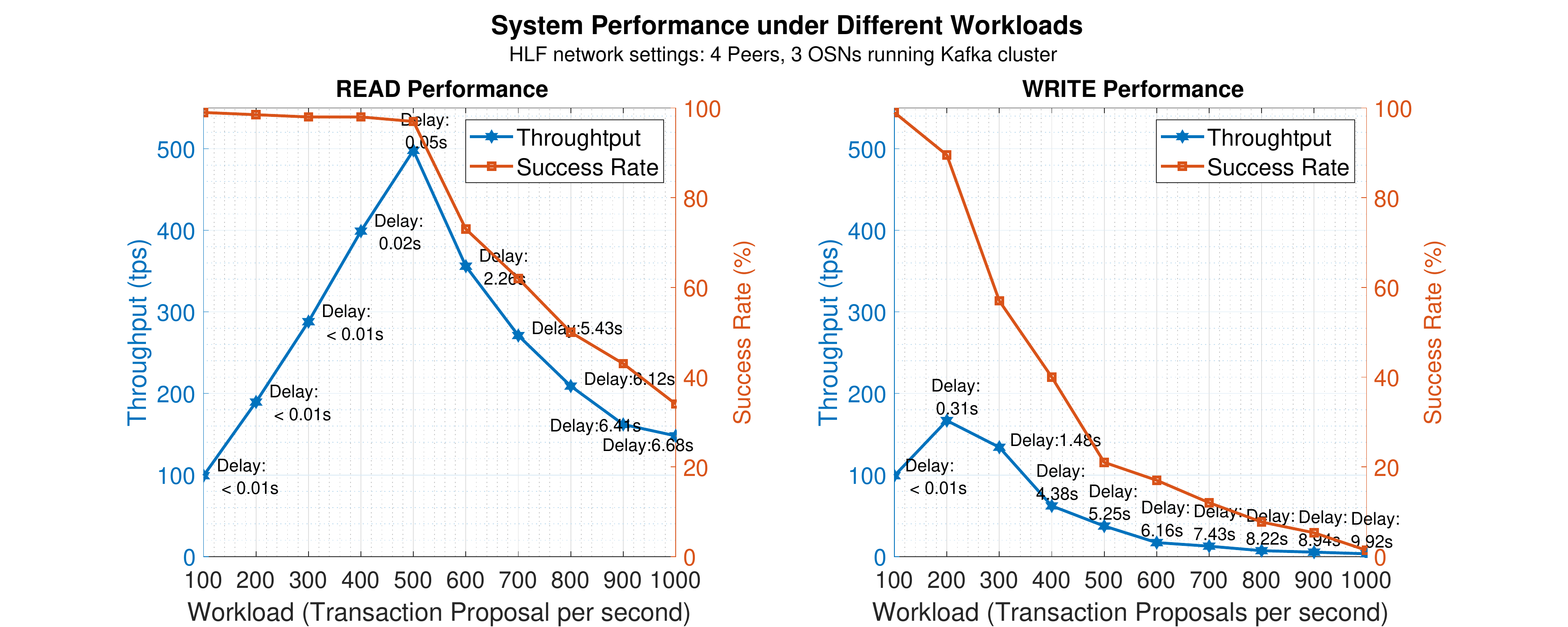}
	\caption{Performance of $READ$ and $WRITE$ from/to distributed ledgers in the HLF-based system under different workloads}
	\label{fig7}
\end{figure*}

As the proposed platform is expected to serve a large number of clients accessing data simultaneously, performance, and scalability of the platform is necessarily evaluated. The core technology leaders in BC such as Bitcoin, Ethereum Enterprise Alliance and Hyperledger Foundation have demonstrated promising technology advancements of both performance and scalability. However, at the moment, public BCs can only achieve limited throughput (e.g., Bitcoin gets $7$ transactions per second ($tps$) with $Blocktime$ is around 10 minutes whereas Ethereum reaches around $15$ $tps$ with 15-second $Blocktime$ \footnote{https://bitinfocharts.com/}). In permissioned BCs, additional permission control ensures that a majority of nodes are trusted; as well as all identities of the participants in the network are known. This allows the use of BFT-variant consensus in the BC platforms, theoretically resulting in higher throughput. For instance, FabricCoin deployed on top of the HLF framework can achieve about $3,500$ $tps$ at a second latency \cite{ref50}. However, scalability incurs as a critical issue for a permissioned BC framework, especially frameworks with the pBFT-variant consensus mechanisms.

\subsubsection{Hyperledger Caliper Performance Benchmark}
For our performance evaluation, we use a new evaluation tool developed by the Hyperledger Foundation called Caliper\footnote{https://hyperledger.github.io/caliper/}, which is a performance benchmark framework for various BC frameworks including HLF, Hyperledger Sawtooth and Ethereum. Caliper is equipped with adaptors implementing interfaces for interacting with HLF systems version $1.x$ using either HLF native SDK or a RESTful API. To integrate with our existing HLF profile management system, we have programmed our adaptors using Fabric Client SDK (NodeJS version) to interact with the BC network and to invoke the two chaincodes $3A\_cc$ and $log\_cc$. On top of the adaptation layer is a benchmark layer implementing predefined use-cases in the form of $YAML$ configuration files. We have written various use-cases for the performance benchmark following these configurations:
\begin{itemize}
	\item $READ$ the ledgers (e.g., invoke \textit{policy\_check} function) and $WRITE$ the the ledgers (e.g., invoke the $GrantConsent$ and $RevokeConsent$ functions to update the $ACL$ policy).
	\item Different HLF network settings in which the number of peer nodes are varied from $4$ to $32$.
	\item Different workloads to the system by generating a number of transaction proposals per second to the system. In each network setting, the workload is from $100 tps$ to $1000 tps$.
\end{itemize}

\subsubsection{Results and Analysis}
There are four metrics in the Caliper benchmarking results, namely \textit{(1)} Success Rate, \textit{(2)} Throughput, \textit{(3)} Latency, and \textit{(4)} Resource Consumption. These metrics are counted from the time a transaction submitted by a client until it is processed and is written on a distributed ledger. Fig. \ref{fig7} interprets our system performance under different number of workloads, from 100tps to 1000tps. The HLF network setting includes 4 peer nodes and 3 OSNs running Kafka cluster for crash-fault tolerance consensus. In this benchmark, $1000$ clients are popularised that generates transaction proposals (including both $READ$ and $WRITE$ a distributed ledger) to our system. As can be seen in the figure, throughput of $READ$ transactions can reach highest to $492$tps at 500-tps workload whereas $WRITE$ transactions only reach to $167$ tps at 300-tps workload with highest success rate (more than 95\%) and with less than 1-second latency. Compared to $READ$ transactions, $WRITE$ transactions require more processes from OSNs to chronologically order the transactions, create a new block, and broadcast it to all peers in the network to update a distributed ledger; that is why $WRITE$ transactions get lower throughput, lower success rate, and higher latency. After these peaks, the throughputs and the success rates of both $READ$ and $WRITE$ transactions dramatically decrease. For instance, at 1000-tps workload, the throughputs and success rates drop to about $34.5$ tps and 30\%; and $3.4$ tps and $1.4$\% for $READ$ and $WRITE$ transactions, respectively. The average latency is significantly risen from less than $1$ second to $6.68$ seconds ($READ$ transaction) and $9.92$ seconds ($WRITE$ transactions) as higher workload is generated.

Generally, the reason that the system cannot handle high workload due to local processing bottleneck as transactions are queued at endorsing peer's buffer and OSNs' buffers (for $WRITE$ transactions) to be processed. The HLF procedure requires that a transaction has to obtain enough proposal responses from endorsing peers, thus, if an endorsing peer processes the transaction lower than others, the transaction is delayed accordingly. Particularly for $WRITE$ transactions which require more processes for ordering service as such, all $WRITE$ transactions need to be buffered and processed at the 3 OSNs running Kafka cluster. As we observed, OSNs are always busy that the docker container can consume to 88\% CPU-load on average.

\begin{figure}[ht]
\centering
\captionsetup{justification=centering}
	\includegraphics[width=0.48\textwidth]{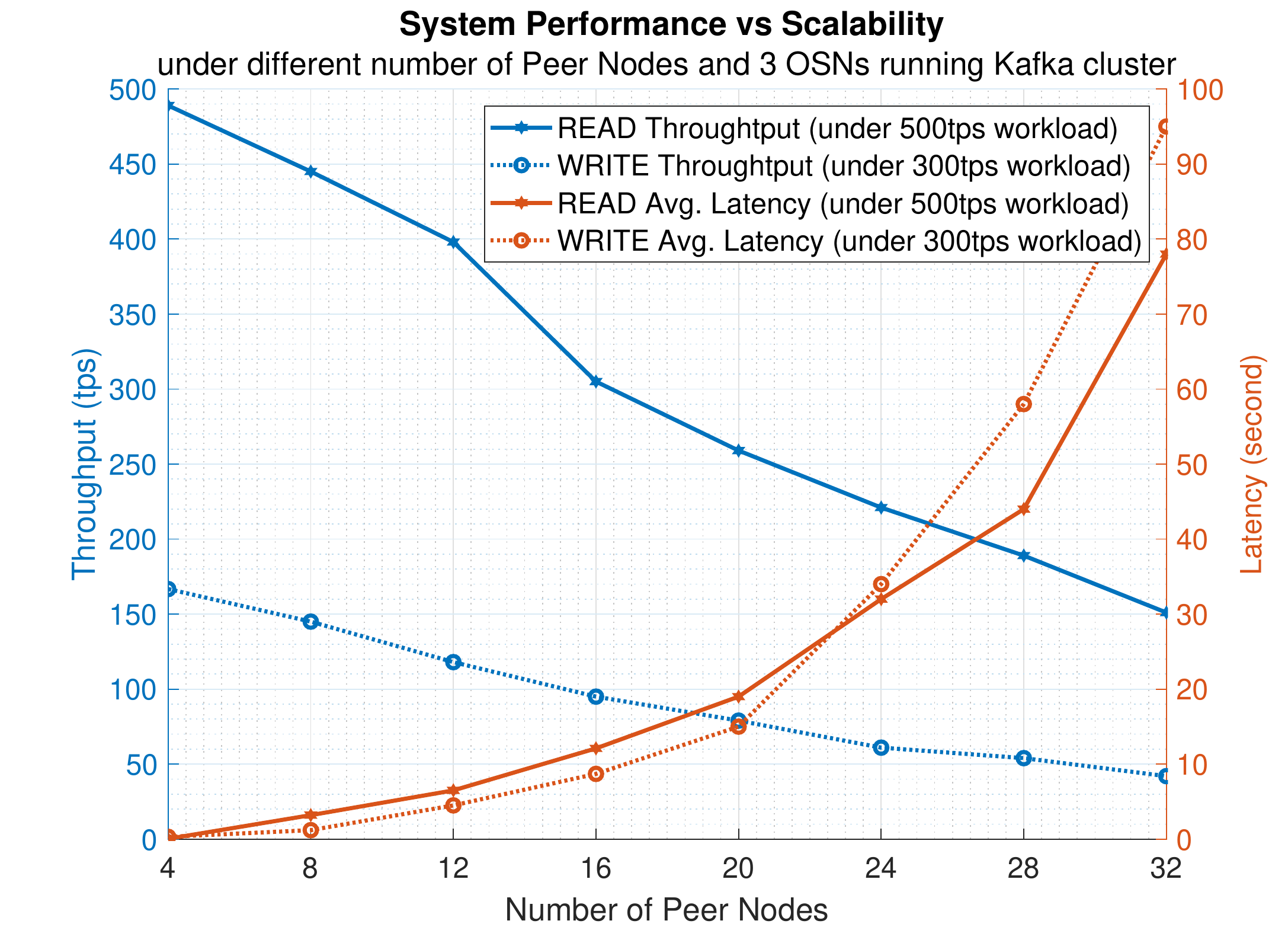}
	\caption{System performance vs scalability under different number of peer nodes}
	\label{fig8}
\end{figure}

Fig. \ref{fig8} illustrates the performance vs. scalability of our proposed system when the number of peer nodes increasing while the ordering service remains the same with $3$ OSNs running Kafka cluster. In this performance benchmark test, $READ$ and $WRITE$ transactions are set under $500$ tps and $300$ tps workload, respectively. More peer nodes mean more overhead messages exchanged between nodes, and the wait for endorsement messages before broadcasting a transaction response to the OSNs to create a new block and to update a distributed ledger. That is why the throughput decreases and the latency increases for both $READ$ and $WRITE$ transactions. As depicted in Fig. \ref{fig8}, the proposed HLF-based profile management system fails to support high performance and scalability since the throughput significantly decreases and the latency dramatically increases when the BC network scales up (e.g., at 32 peer nodes, throughputs are $151$ tps and $42$ tps and latencies are $78$s and $95$s for $READ$ and $WRITE$ transactions, respectively). Fortunately, HLF allows us to partition a BC network in which only a subset of peer nodes are permitted to endorse a particular chain-code. This will reduce the number of messages exchanged across the network as well as reduce the waiting time for endorsement messages from endorsing peers. As a trade-off, decentralisation is partly sacrificed and the system is more sensitive to 51\% and selfish mining attacks \cite{ref_add_06}.

%% file: 7Conclusion.tex
\section{Conclusion and the road ahead}
In this article, a design concept for a GDPR-compliant BC-based personal data management platform is proposed. Following the guidelines from the design concept including system architecture, ledger data models, and SC functionalities, a BC-based platform is implemented on top of the HLF framework. The platform interplays among an honest RS, a social networking SP, DPs, and DSs ensuring that all processing activities over profile data stored in the RS are compliant with the GDPR. The feasibility and effectiveness of the design concept are, therefore, successfully demonstrated.

For future work, we will deploy the design concept in a public BC (e.g., Ethereum) with an RS using distributed storage (e.g., IPFS, BigchainDB or Storj). In this regard, the RS is not trustworthy as some storage nodes might be malicious. Thus, more mechanisms need to be implemented to resolve the lack of a trusted centralised RS. As a reward, the system is truly decentralised. Another work is to develop a fine-grain expressive data usage policy using a policy language instead of a simple ACL as in the demonstration. A policy generator deployed in SCs that autonomously acquires data usage policy depending on specific contexts is also a promising research direction. Additionally, pricing and incentive models for the cost of data storage and BC operations should be carried out to finalise a complete system.

As the processing of personal data refers to CRUD operations – which is under the mindset of data storage, an ambitious research direction is to provide computational capability on a BC network [32]. This means an SP directly runs computation on the network and obtain results using secure Multi-Party Computation (MPC)\footnote{https://en.wikipedia.org/wiki/Secure\_multi-party\_computation}. This approach is much securer as the SP does not directly observe raw data. We believe our work acts as a catalyst to open a variety of research directions regarding the use of BC and SCs in decentralised authorisation and access control, which plays a crucial role in digital assets management, particularly in personal data regulations.

%% file: main.bbl
\begin{thebibliography}{10}
\providecommand{\url}[1]{#1}
\csname url@samestyle\endcsname
\providecommand{\newblock}{\relax}
\providecommand{\bibinfo}[2]{#2}
\providecommand{\BIBentrySTDinterwordspacing}{\spaceskip=0pt\relax}
\providecommand{\BIBentryALTinterwordstretchfactor}{4}
\providecommand{\BIBentryALTinterwordspacing}{\spaceskip=\fontdimen2\font plus
\BIBentryALTinterwordstretchfactor\fontdimen3\font minus
  \fontdimen4\font\relax}
\providecommand{\BIBforeignlanguage}[2]{{%
\expandafter\ifx\csname l@#1\endcsname\relax
\typeout{** WARNING: IEEEtran.bst: No hyphenation pattern has been}%
\typeout{** loaded for the language `#1'. Using the pattern for}%
\typeout{** the default language instead.}%
\else
\language=\csname l@#1\endcsname
\fi
#2}}
\providecommand{\BIBdecl}{\relax}
\BIBdecl

\bibitem{ref24}
H.~Shafagh, L.~Burkhalter, A.~Hithnawi, and S.~Duquennoy, ``Towards
  blockchain-based auditable storage and sharing of iot data,'' in
  \emph{Proceedings of the 2017 on Cloud Computing Security Workshop}.\hskip
  1em plus 0.5em minus 0.4em\relax ACM, 2017, pp. 45--50.

\bibitem{ref25}
R.~Li, T.~Song, B.~Mei, H.~Li, X.~Cheng, and L.~Sun, ``Blockchain for
  large-scale internet of things data storage and protection,'' \emph{IEEE
  Transactions on Services Computing}, 2018.

\bibitem{ref30}
S.~Wang, Y.~Zhang, and Y.~Zhang, ``A blockchain-based framework for data
  sharing with fine-grained access control in decentralized storage systems,''
  \emph{IEEE Access}, vol.~6, pp. 38\,437--38\,450, 2018.

\bibitem{ref31}
J.~Benet, ``Ipfs-content addressed, versioned, p2p file system,'' \emph{arXiv
  preprint arXiv:1407.3561}, 2014.

\bibitem{ref32}
G.~Zyskind, O.~Nathan \emph{et~al.}, ``Decentralizing privacy: Using blockchain
  to protect personal data,'' in \emph{2015 IEEE Security and Privacy
  Workshops}.\hskip 1em plus 0.5em minus 0.4em\relax IEEE, 2015, pp. 180--184.

\bibitem{ref33}
L.~A. Linn and M.~B. Koo, ``Blockchain for health data and its potential use in
  health it and health care related research,'' in \emph{ONC/NIST Use of
  Blockchain for Healthcare and Research Workshop. Gaithersburg, Maryland,
  United States: ONC/NIST}, 2016.

\bibitem{ref34}
A.~Azaria, A.~Ekblaw, T.~Vieira, and A.~Lippman, ``Medrec: Using blockchain for
  medical data access and permission management,'' in \emph{2016 2nd
  International Conference on Open and Big Data (OBD)}.\hskip 1em plus 0.5em
  minus 0.4em\relax IEEE, 2016, pp. 25--30.

\bibitem{ref35}
M.~J.~M. Chowdhury, A.~Colman, M.~A. Kabir, J.~Han, and P.~Sarda, ``Blockchain
  as a notarization service for data sharing with personal data store,'' in
  \emph{2018 17th IEEE International Conference On Trust, Security And Privacy
  In Computing And Communications(TrustCom)}.\hskip 1em plus 0.5em minus
  0.4em\relax IEEE, 2018, pp. 1330--1335.

\bibitem{ref36}
R.~Neisse, G.~Steri, and I.~Nai-Fovino, ``A blockchain-based approach for data
  accountability and provenance tracking,'' in \emph{Proceedings of the 12th
  International Conference on Availability, Reliability and Security}.\hskip
  1em plus 0.5em minus 0.4em\relax ACM, 2017, p.~14.

\bibitem{ref37}
B.~Faber, G.~C. Michelet, N.~Weidmann, R.~R. Mukkamala, and R.~Vatrapu,
  ``Bpdims: A blockchain-based personal data and identity management system,''
  in \emph{Proceedings of the 52nd Hawaii International Conference on System
  Sciences}, 2019.

\bibitem{ref38}
C.~Wirth and M.~Kolain, ``Privacy by blockchain design: a blockchain-enabled
  gdpr-compliant approach for handling personal data,'' in \emph{Proceedings of
  1st ERCIM Blockchain Workshop 2018}.\hskip 1em plus 0.5em minus 0.4em\relax
  European Society for Socially Embedded Technologies (EUSSET), 2018.

\bibitem{ref01}
M.~Walport \emph{et~al.}, ``Distributed ledger technology: Beyond blockchain,''
  \emph{UK Government Office for Science}, vol.~1, 2016.

\bibitem{ref_add_02}
P.~Voigt and A.~Von~dem Bussche, ``The eu general data protection regulation
  (gdpr),'' \emph{A Practical Guide, 1st Ed., Cham: Springer International
  Publishing}, 2017.

\bibitem{ref_add_01}
I.~P. Team, \emph{EU general data protection regulation (GDPR): an
  implementation and compliance guide}.\hskip 1em plus 0.5em minus 0.4em\relax
  IT Governance Ltd, 2017.

\bibitem{ref02}
S.~Nakamoto, ``Bitcoin: A peer-to-peer electronic cash system,'' 2008.

\bibitem{ref03}
M.~Crosby, P.~Pattanayak, S.~Verma, V.~Kalyanaraman \emph{et~al.}, ``Blockchain
  technology: Beyond bitcoin,'' \emph{Applied Innovation}, vol.~2, no. 6-10,
  p.~71, 2016.

\bibitem{ref04}
F.~Tschorsch and B.~Scheuermann, ``Bitcoin and beyond: A technical survey on
  decentralized digital currencies,'' \emph{IEEE Communications Surveys \&
  Tutorials}, vol.~18, no.~3, pp. 2084--2123, 2016.

\bibitem{ref05}
N.~B. Truong, T.-W. Um, B.~Zhou, and G.~M. Lee, ``Strengthening the
  blockchain-based internet of value with trust,'' in \emph{2018 IEEE
  International Conference on Communications (ICC)}.\hskip 1em plus 0.5em minus
  0.4em\relax IEEE, 2018, pp. 1--7.

\bibitem{ref06}
V.~Gramoli, ``From blockchain consensus back to byzantine consensus,''
  \emph{Future Generation Computer Systems}, 2017.

\bibitem{ref07}
W.~Wang, D.~T. Hoang, P.~Hu, Z.~Xiong, D.~Niyato, P.~Wang, Y.~Wen, and D.~I.
  Kim, ``A survey on consensus mechanisms and mining strategy management in
  blockchain networks,'' \emph{IEEE Access}, 2019.

\bibitem{ref08}
A.~Gervais, G.~O. Karame, K.~W{\"u}st, V.~Glykantzis, H.~Ritzdorf, and
  S.~Capkun, ``On the security and performance of proof of work blockchains,''
  in \emph{Proceedings of the 2016 ACM SIGSAC conference on computer and
  communications security}.\hskip 1em plus 0.5em minus 0.4em\relax ACM, 2016,
  pp. 3--16.

\bibitem{ref09}
A.~Kiayias, A.~Russell, B.~David, and R.~Oliynykov, ``Ouroboros: A provably
  secure proof-of-stake blockchain protocol,'' in \emph{Annual International
  Cryptology Conference}.\hskip 1em plus 0.5em minus 0.4em\relax Springer,
  2017, pp. 357--388.

\bibitem{ref10}
I.~Bentov, C.~Lee, A.~Mizrahi, and M.~Rosenfeld, ``Proof of activity: Extending
  bitcoin's proof of work via proof of stake.'' \emph{IACR Cryptology ePrint
  Archive}, vol. 2014, p. 452, 2014.

\bibitem{ref11}
A.~Miller and J.~J. LaViola~Jr, ``Anonymous byzantine consensus from
  moderately-hard puzzles: A model for bitcoin,'' \emph{Available online:
  http://nakamotoinstitute.org/research/anonymous-byzantine-consensus}, 2014.

\bibitem{ref12}
Y.~Gilad, R.~Hemo, S.~Micali, G.~Vlachos, and N.~Zeldovich, ``Algorand: Scaling
  byzantine agreements for cryptocurrencies,'' in \emph{Proceedings of the 26th
  Symposium on Operating Systems Principles}.\hskip 1em plus 0.5em minus
  0.4em\relax ACM, 2017, pp. 51--68.

\bibitem{ref13}
V.~Buterin, ``White paper: A next-generation smart contract and decentralized
  application platform,'' \emph{April. https://www. ethereum.
  org/pdfs/EthereumWhitePaper. pdf}, 2014.

\bibitem{ref14}
A.~Kosba, A.~Miller, E.~Shi, Z.~Wen, and C.~Papamanthou, ``Hawk: The blockchain
  model of cryptography and privacy-preserving smart contracts,'' in \emph{2016
  IEEE symposium on security and privacy (SP)}.\hskip 1em plus 0.5em minus
  0.4em\relax IEEE, 2016, pp. 839--858.

\bibitem{ref15}
G.~Wood, ``Ethereum: A secure decentralised generalised transaction ledger,''
  \emph{Ethereum project yellow paper}, vol. 151, pp. 1--32, 2014.

\bibitem{ref16}
C.~Cachin, ``Architecture of the hyperledger blockchain fabric,'' in
  \emph{Workshop on distributed cryptocurrencies and consensus ledgers}, vol.
  310, 2016.

\bibitem{ref17}
\BIBentryALTinterwordspacing
H.~A. WC. (2018) Hyperledger architecture-volume ii-smart contracts. [Online].
  Available:
  \url{https://www.hyperledger.org/wp-content/uploads/2018/04/Hyperledger_Arch_WG_Paper_2_SmartContracts.pdf}
\BIBentrySTDinterwordspacing

\bibitem{ref18}
K.~Markus and G.~Chung, ``Blockchain in logistics,'' DHL Trend Research,
  Germany, 2018.

\bibitem{ref19}
F.~Tian, ``An agri-food supply chain traceability system for china based on
  rfid \& blockchain technology,'' in \emph{2016 13th international conference
  on service systems and service management (ICSSSM)}.\hskip 1em plus 0.5em
  minus 0.4em\relax IEEE, 2016, pp. 1--6.

\bibitem{ref20}
N.~Hackius and M.~Petersen, ``Blockchain in logistics and supply chain: trick
  or treat?'' in \emph{Proceedings of the Hamburg International Conference of
  Logistics (HICL)}.\hskip 1em plus 0.5em minus 0.4em\relax epubli, 2017, pp.
  3--18.

\bibitem{ref21}
X.~Liang, S.~Shetty, D.~Tosh, C.~Kamhoua, K.~Kwiat, and L.~Njilla, ``Provchain:
  A blockchain-based data provenance architecture in cloud environment with
  enhanced privacy and availability,'' in \emph{Proceedings of the 17th
  IEEE/ACM international symposium on cluster, cloud and grid computing}.\hskip
  1em plus 0.5em minus 0.4em\relax IEEE Press, 2017, pp. 468--477.

\bibitem{ref22}
M.~Ali, J.~Nelson, R.~Shea, and M.~J. Freedman, ``Blockstack: A global naming
  and storage system secured by blockchains,'' in \emph{Annual Technical
  Conference (USENIX/ATC-16)}, 2016, pp. 181--194.

\bibitem{ref23}
H.~A. Kalodner, M.~Carlsten, P.~Ellenbogen, J.~Bonneau, and A.~Narayanan, ``An
  empirical study of namecoin and lessons for decentralized namespace design.''
  in \emph{WEIS}.\hskip 1em plus 0.5em minus 0.4em\relax Citeseer, 2015.

\bibitem{ref26}
T.~McConaghy, R.~Marques, A.~M{\"u}ller, D.~De~Jonghe, T.~McConaghy,
  G.~McMullen, R.~Henderson, S.~Bellemare, and A.~Granzotto, ``Bigchaindb: a
  scalable blockchain database,'' \emph{White paper, BigChainDB}, 2016.

\bibitem{ref27}
J.-H. Lee, ``Bidaas: Blockchain based id as a service,'' \emph{IEEE Access},
  vol.~6, pp. 2274--2278, 2018.

\bibitem{ref28}
Z.~Chen, S.~Chen, H.~Xu, and B.~Hu, ``A security authentication scheme of 5g
  ultra-dense network based on block chain,'' \emph{IEEE Access}, vol.~6, pp.
  55\,372--55\,379, 2018.

\bibitem{ref29}
O.~Novo, ``Blockchain meets iot: An architecture for scalable access management
  in iot,'' \emph{IEEE Internet of Things Journal}, vol.~5, no.~2, pp.
  1184--1195, 2018.

\bibitem{ref39}
D.~Hardt, ``The oauth 2.0 authorization framework,'' IETF, Tech. Rep., 2012.

\bibitem{ref40}
T.~Lodderstedt, M.~McGloin, and P.~Hunt, ``Oauth 2.0 threat model and security
  considerations,'' IETF, Tech. Rep., 2013.

\bibitem{ref41}
R.~Neisse, G.~Steri, I.~N. Fovino, and G.~Baldini, ``Seckit: a model-based
  security toolkit for the internet of things,'' \emph{computers \& security},
  vol.~54, pp. 60--76, 2015.

\bibitem{ref42}
M.~Berberich and M.~Steiner, ``Blockchain technology and the gdpr-how to
  reconcile privacy and distributed ledgers,'' \emph{European Data Protection
  Law Review}, vol.~2, no. 422, 2016.

\bibitem{ref43}
S.~Wilkinson, T.~Boshevski, J.~Brandoff, and V.~Buterin, ``Storj a peer-to-peer
  cloud storage network,'' \emph{CiteSeer}, 2014.

\bibitem{ref_add_03}
D.~Johnson, A.~Menezes, and S.~Vanstone, ``The elliptic curve digital signature
  algorithm (ecdsa),'' \emph{International journal of information security},
  vol.~1, no.~1, pp. 36--63, 2001.

\bibitem{ref44}
D.~Schwartz, N.~Youngs, A.~Britto \emph{et~al.}, ``The ripple protocol
  consensus algorithm,'' \emph{Ripple Labs Inc White Paper}, vol.~5, 2014.

\bibitem{ref45}
S.~Liu, P.~Viotti, C.~Cachin, V.~Qu{\'e}ma, and M.~Vukoli{\'c}, ``$\{$XFT$\}$:
  Practical fault tolerance beyond crashes,'' in \emph{12th $\{$USENIX$\}$
  Symposium on Operating Systems Design and Implementation ($\{$OSDI$\}$ 16)},
  2016, pp. 485--500.

\bibitem{ref46}
\BIBentryALTinterwordspacing
C.~R. Meijer. (2018) Blockchain versus gdpr and who should adjust most.
  [Online]. Available:
  \url{https://www.finextra.com/blogposting/16102/blockchain-versus-gdpr-and-who-should-adjust-most}
\BIBentrySTDinterwordspacing

\bibitem{ref47}
H.~Zhao, P.~Bai, Y.~Peng, and R.~Xu, ``Efficient key management scheme for
  health blockchain,'' \emph{CAAI Transactions on Intelligence Technology},
  vol.~3, no.~2, pp. 114--118, 2018.

\bibitem{ref48}
L.~Luu, D.-H. Chu, H.~Olickel, P.~Saxena, and A.~Hobor, ``Making smart
  contracts smarter,'' in \emph{Proceedings of the 2016 ACM SIGSAC Conference
  on Computer and Communications Security}.\hskip 1em plus 0.5em minus
  0.4em\relax ACM, 2016, pp. 254--269.

\bibitem{ref49}
S.~Meiklejohn, M.~Pomarole, G.~Jordan, K.~Levchenko, D.~McCoy, G.~M. Voelker,
  and S.~Savage, ``A fistful of bitcoins: characterizing payments among men
  with no names,'' in \emph{Proceedings of the 2013 conference on Internet
  measurement conference}.\hskip 1em plus 0.5em minus 0.4em\relax ACM, 2013,
  pp. 127--140.

\bibitem{ref50}
E.~Androulaki, A.~Barger, V.~Bortnikov, C.~Cachin, K.~Christidis, A.~De~Caro,
  D.~Enyeart, C.~Ferris, G.~Laventman, Y.~Manevich \emph{et~al.}, ``Hyperledger
  fabric: a distributed operating system for permissioned blockchains,'' in
  \emph{Proceedings of the Thirteenth EuroSys Conference}.\hskip 1em plus 0.5em
  minus 0.4em\relax ACM, 2018, p.~30.

\bibitem{ref_add_04}
C.~Gentry \emph{et~al.}, ``Fully homomorphic encryption using ideal lattices.''
  in \emph{Stoc}, vol.~9, no. 2009, 2009, pp. 169--178.

\bibitem{ref_add_05}
V.~Goyal, O.~Pandey, A.~Sahai, and B.~Waters, ``Attribute-based encryption for
  fine-grained access control of encrypted data,'' in \emph{Proceedings of the
  13th ACM conference on Computer and communications security}.\hskip 1em plus
  0.5em minus 0.4em\relax Acm, 2006, pp. 89--98.

\bibitem{ref_add_06}
I.-C. Lin and T.-C. Liao, ``A survey of blockchain security issues and
  challenges.'' \emph{IJ Network Security}, vol.~19, no.~5, pp. 653--659, 2017.

\end{thebibliography}
